
\magnification\magstep1

\openup 1\jot

\input mssymb
\def\hbar{\mathchar '26\mkern -9muh}

\catcode`@=11
\def\eqaltxt#1{\displ@y \tabskip 0pt
  \halign to\displaywidth {%
    \rlap{$##$}\tabskip\centering
    &\hfil$\@lign\displaystyle{##}$\tabskip\z@skip
    &$\@lign\displaystyle{{}##}$\hfil\tabskip\centering
    &\llap{$\@lign##$}\tabskip\z@skip\crcr
    #1\crcr}}
\def\eqallft#1{\displ@y \tabskip 0pt
  \halign to\displaywidth {%
    $\@lign\displaystyle {##}$\tabskip\z@skip
    &$\@lign\displaystyle{{}##}$\hfil\crcr
    #1\crcr}}
\catcode`@=12 

\def\half{{\textstyle {1 \over 2}}}

\def\pmb#1{\setbox0=\hbox{#1}  \kern-.025em\copy0\kern-\wd0
  \kern0.05em\copy0\kern-\wd0  \kern-.025em\raise.0433em\box0 }
\def\pmbh#1{\setbox0=\hbox{#1} \kern-.12em\copy0\kern-\wd0
	    \kern.12em\copy0\kern-\wd0\box0}
\def\sqr#1#2{{\vcenter{\vbox{\hrule height.#2pt
      \hbox{\vrule width.#2pt height#1pt \kern#1pt
	 \vrule width.#2pt}
      \hrule height.#2pt}}}}

\def\rchi{{\raise 2pt \hbox {$\chi$}}}
\def\rga{{\raise 2pt \hbox {$\gamma$}}}
\def\rg{{\raise 2 pt \hbox {$g$}}}

\def\susy{supersymmetry}
\def\({\left(}
\def\){\right)}
\def\<{\left\langle}
\def\>{\right\rangle}

\def\[{\left[}
\def\]{\right]}
\let\text=\hbox
\def\pt{\partial}
\def\eps{\epsilon}
\def\kap{\kappa}
\def\al{\alpha}

\def\om{\omega}
\def\ol{\overline}

\def\de{\delta}
\def\lam{\lambda}

\def\sig{\sigma}

\def\ti{\tilde}

\def\cH{{\cal H}}

\def\cD{{\cal D}}

\def\Lam{\Lambda}

\def\wti{\widetilde}
\def\Ga{\Gamma}

\def\A{\hbox{$ A\kern -5.5pt / \kern +5.5pt$}}
\def\B{\hbox{$ B\kern -6.5pt / \kern +.5pt$}}
\def\C{\hbox{$ C\kern -6.5pt / \kern +.5pt$}}
\def\D{\hbox{$ D\kern -6.5pt / \kern +.5pt$}}
\def\E{\hbox{$ E\kern -6.5pt / \kern +.5pt$}}
\def\F{\hbox{$ F\kern -6.5pt / \kern +.5pt$}}
\def\G{\hbox{$ G\kern -6.5pt / \kern +.5pt$}}
\def\H{\hbox{$ H\kern -7.5pt / \kern +.5pt$}}
\def\I{\hbox{$ {\bf I}\kern -4.5pt / \kern +.5pt$}}
\def\Z{\hbox{$ Z\kern -6.5pt / \kern +.5pt$}}

\hfuzz 6pt

\catcode`@=12 
\rightline {\bf DAMTP R95/21}

\vskip .2 true in
\centerline {\bf Quantization of the Bianchi type-IX  model in N=1 Supergravity
 }
\centerline {\bf in the Presence of
 Supermatter}
\vskip .1 true in
\centerline {
{\rm P.V. Moniz}\footnote*{{\rm e-mail address: prlvm10@amtp.cam.ac.uk}}}
\vskip .1 true in
\centerline {Department of Applied Mathematics and Theoretical Physics}
\centerline {University of Cambridge}
\centerline { Silver Street, Cambridge}
\centerline {CB3 9EW, UK }
\vskip .2 true in
\centerline {\bf ABSTRACT}
\vskip .1 true in

The general theory of N=1 supergravity with supermatter is applied
to a Bianchi type IX diagonal model. The supermatter is constituted
by a complex scalar field and its spin-$1\over 2$ fermionic partners.
The K\"ahler geometry is chosen to be a two-dimensional flat one.
The Lorentz invariant Ansatz for the wave function of the
universe is taken to be as simple as possible in order to obtain
{\it new} solutions.
The
set of differential equations derived from the
quantum constraints are analysed in two   different cases:
 if  the supermatter terms include an analytical
potential or not. In the latter the wave function
is found to have a simple form.

\vfill
\magnification\magstep1
\centerline {PACS numbers: 04.60.+ $n$, 04.65.+ $e$, 98.80. $Hw$ }
\vskip 6 pt
\noindent

\vfill
\eject
\noindent

\magnification\magstep1

\openup 0.7\jot

\input mssymb
\def\hbar{\mathchar '26\mkern -9muh}

\catcode`@=11
\def\eqaltxt#1{\displ@y \tabskip 0pt
  \halign to\displaywidth {%
    \rlap{$##$}\tabskip\centering
    &\hfil$\@lign\displaystyle{##}$\tabskip\z@skip
    &$\@lign\displaystyle{{}##}$\hfil\tabskip\centering
    &\llap{$\@lign##$}\tabskip\z@skip\crcr
    #1\crcr}}
\def\eqallft#1{\displ@y \tabskip 0pt
  \halign to\displaywidth {%
    $\@lign\displaystyle {##}$\tabskip\z@skip
    &$\@lign\displaystyle{{}##}$\hfil\crcr
    #1\crcr}}
\catcode`@=12 

\def\half{{\textstyle {1 \over 2}}}

\def\pmb#1{\setbox0=\hbox{#1}  \kern-.025em\copy0\kern-\wd0
  \kern0.05em\copy0\kern-\wd0  \kern-.025em\raise.0433em\box0 }

\def\pmbh#1{\setbox0=\hbox{#1} \kern-.12em\copy0\kern-\wd0
	    \kern.12em\copy0\kern-\wd0\box0}
\def\sqr#1#2{{\vcenter{\vbox{\hrule height.#2pt
      \hbox{\vrule width.#2pt height#1pt \kern#1pt
	 \vrule width.#2pt}
      \hrule height.#2pt}}}}

\def\rchi{{\raise 2pt \hbox {$\chi$}}}
\def\rga{{\raise 2pt \hbox {$\gamma$}}}
\def\rrho{{\raise 2pt \hbox {$\rho$}}}

\def\({\left(}
\def\){\right)}
\def\<{\left\langle}
\def\>{\right\rangle}

\def\[{\left[}
\def\]{\right]}
\let\text=\hbox
\def\pt{\partial}
\def\eps{\epsilon}
\def\kap{\kappa}
\def\al{\alpha}

\def\om{\omega}
\def\ol{\overline}

\def\de{\delta}
\def\lam{\lambda}

\def\sig{\sigma}

\def\ti{\tilde}

\def\Lam{\Lambda}

\def\wti{\widetilde}
\def\Ga{\Gamma}

\def\cH{{\cal H}}


{\bf I. Introduction}

The subjects of   supersymmetric
quantum gravity and cosmology have achieved a number
of interesting results and conclusions during the last ten years or so.
Several approaches
may be found  in the literature, namely the triad ADM
canonical formulation [1--26,62,63], the $\sigma-$model
supersymmetric extension in quantum cosmology [27--31]
and another approach based on Ashtekar variables
 [32--40].
Quite recently, important contributions have also
been made [22,23,24], which  point towards a well desired
revival of the field.
A  review on these  subjects is currently in preparation [41].

The  canonical quantization framework of N=1 (pure) supergravity was
presented in ref. [1],
following [42,43]. It was pointed out that it
would be  sufficient,
in finding a physical state, to solve
the Lorentz and supersymmetry constraints of the theory  because the
algebra of constraints of the theory leads to anti-commutation relations
implying   that   a
physical wave functional $ \Psi $ will also obey
 the Hamiltonian constraints [1,42]. The factor ordering of the  Hamiltonian
constraints
is determined by the anti-commutation relations of the supersymmetry
constraints.
Namely, it will  depend on how we order
the fermionic derivatives in the supersymmetry constraints,
 which are enforced by the ordering
in the spinorial form of the gravitational momentum. More precisely,
such factor ordering
implies that supersymmetry constraints should describe
 the left and right handed supersymmetry transformations (cf. ref.[1])
(when considering reduced minisuperspace models, different factor ordering
have been chosen
[10-12,18,25]).

Using the triad ADM
canonical formulation, Bianchi models in pure $N=1$ supergravity have been
studied in ref [2-8,22,23,24] . The quantum states may be described by a
 wave function of the form  $\Psi( e_{AA'i}
, \psi_{Ai})$ where $e_{AA'i}$ and $\psi_{Ai}$ denote, respectively, the
two-component spinor form of
the tetrad and the spin-$3 \over 2$ gravitino field.
 The wave function may be then expanded in even powers of $\psi_{Ai}$,
symbolically represented by $\psi^{0}, \psi^{2}, \psi^{4}$ up to $\psi^{6}$
 because of the anti-commutations relations of the six spatial components of
 the gravitino fields (see  ref.[6,7,15,16,24] for more details).

The  analysis of locally
supersymetric Bianchi class A models also began  to face some unexpected
difficulties.
Supersymmetry (as well as other considerations) forbids mini-superspace
models of class $ B $. Firstly,
models without supermatter had (simple) solutions
{\bf only} in the empty $\psi^0$ (bosonic) and fermionic filled $\psi^6$
sectors. More precisely, the physical states in these sectors
were, respectively, given by (cf. ref. [6-8,13,16,22,23,24])
$$ \psi^0 \rightarrow e^{ m^{pq}h_{pq}}, \eqno(1.1) $$
$$ \psi^6 \rightarrow h e^{-m^{pq}h_{pq}} \Pi_{i} (\psi^{A}_i)^2.
\eqno(1.2) $$
Here $h$ is the determinant of the 3-metric $h_{pq}$ and
$m^{pq}$ is defined from the
relation
$$ d w^p = {1\over 2} m^{pq} h^{-\half} \epsilon_{qrs}
\omega^r \otimes \omega^s, \eqno(1.3)$$
where $\omega^r$ are basis of left-invariant 1-forms on the
space-like hypersurface of homogeneity; the constant symmetric
matrix $m^{pq}$ is fixed by the chosen Bianchi type (cf. e.g.,
[44]). Secondly, these states could be interpreted either as Hartle-Hawking
no-boundary solutions [45] or
wormhole states [46]. However, one
could {\it not} found both of them in the same
spectrum of solutions. According to  different homogeneity conditions for
the gravitino field (cf. ref. [8]), we either could find the Hartle-Hawking or
the
wormhole state. In addition,  these solutions in
minisuperspace were shown not to have any counterpart in the
full theory because no states with zero (bosonic) or finite number of
fermions are possible there [19,20]. Finally, when a cosmological constant was
added no
physical states were found [14,15,16,36]
(regarding the  k=1 Friedmann-Robertson-Walker model, where the fermionic
degrees of freedom of the gravitino field
are very restricted,
a bosonic quantum
physical state was found, namely  the Hartle-Hawking solution [45] for a
De Sitter state).  It seemed that the
gravitational and gravitino modes that were allowed to be excited in
 each  supersymmetric Bianchi model contribute in such a way as to
 give only very simple states or even
forbid any physical solutions of the quantum constraints.

All these results seemed difficult to accomodate.
In fact,
doubts were then raised in ref.[23,24]: even though
canonical quantum supergravity has more constraints
than ordinary quantum gravity, it has surely much more
degrees of freedom than gravity\footnote{$^{1}$}{{\sevenrm One should
also stress that the action of the full supergravity theory with boundary terms
[1]
 is not fully invariant under supersymmetry
transformations. The invariance of an action under
the corresponding symmetries of the problem in study is
an obvious desideratum  [47]. In ref. [22]
such invariance for  supersymmetry
transformations was achieved  for the case of Bianchi class A models
using  appropriate extra boundary terms.}}. Hence, why should we experience
problems such as the above expressed, like few or even no physical states in
acceptable physical situations?
The cause for the apparent paradoxical results
mentioned in the previous paragraph
was the use
of an Ansatz too special for the $\psi^2$ and $\psi^4$ fermionic
(middle) sectors in the wave function.
This Ansatz [6-8,13-16,26]
for  Lorentz invariant
fermionic sectors allowed only for two bosonic amplitudes
in each of the $\psi^2$ and $\psi^4$ middle fermionic sectors.
These were constructed only from the Lorentz irreducible
modes of the gravitino field and corresponded to bilinear
and quadratic terms in the gravitino field. However,
there can be  actually  up to 15 such invariants, when we consider
the Lorentz irreducible modes of the gravitational
degrees of freedom as well (see [23,24] for more
details)\footnote{$^{2}$}{{\sevenrm For
the case of a FRW model without supermatter and due
to the restriction of the gravitino field to its
spin-$\half$ mode component, the ``old'' Ansatz for
the wave function remains valid [48].}}.
These 15 Lorentz invariant components for each fermionic
middle sector correspond to  a single one
which satisfy a Wheeler-DeWitt type equation. For a particular
factor ordering, we obtain  the wormhole state in the
bosonic sector and the Hartle-Hawking solution  in the
quartic fermionic sector. The extension of ref. [23,24]
framework to Bianchi models with a cosmological constant
term is currently under way [49].

A subsequent  step would be to  consider
 more general supergravity models involving lower-spin fields.
 One possibility is to take  higher-$N$ gauged supergravity models
[50].
Generically,
  these are technically difficult in the approach used in [1] because
they may  contain a $ \Lam $-term which breaks chirality.
Some results on the canonical quantization of $d=4, N=2$ supergravity
with a non-zero cosmological constant using
Ashtekar variables was presented in [40].
However, the simple $N=2$ supergravity with a {\it global}
$O(2)$ or $SO(3)$ symmetry
[50,51,52] may still prove to be useful
within the ADM approach. In these particular cases,  we do not have
the imposition of a cosmological constant term. An analysis
for the case of a Bianchi I model is
currently in progress [53].
Another possibility
is to  consider the  theory of  $ N = 1 $ supergravity
 coupled to supermatter, and in
particular its supersymmetry constraints (see e.g.,  ref. [54]).
 Its   canonical formulation can be found in refs. [17,18].

 Clearly, a  richer and more
 interesting class of models is given by
coupling  supermatter to $ N = 1 $ supergravity in 4 dimensions.
 In particular,
a dimensional reduction allows  one to
obtain a (1+0)-dimensional theory with N=4 \susy ~
   from (1+3)  dimensional (pure)  N=1 supergravity
[9,10,12].
Although such  minisuperspace constructions
may provide us with further understanding
of some specific aspects which we
hope that will hold in some limit
for the general 4-dimensional theory,
one should also point out their limitations.
The truncation of the inhomogenous modes
constitute a severe restriction, in particulary
in a Friedmann model where the anisotropy
degrees of freedom have been frozen as well. The validity
of the minisuperspace approximation
in locally supersymmetric models is yet an open problem. Nevertheless,
simplified models
like the special case of a FRW universe may allow us to obtain interesting
properties even
if the results may strongly depend on the minisuperspace truncations.
In ref.[9-12]  an Ansatz for the gravitational and
 spin-${3 \over 2}$
fields was introduced in order to reduce
pure $ N = 1 $ supergravity in 4 dimensions to a locally supersymmetric
quantum cosmological model in 1 dimension, assuming a Friedmann $ k = + 1 $
geometry and homogeneity of the spin-${3 \over 2}$ field on the $ S^3 $ spatial
sections. The Hamiltonian structure of the resulting theory was found,
leading to the quantum constraint equations. The general solution to the
quantum constraints is very simple in this case, and the Hartle--Hawking
wave-function can be found.
Following the particular supermatter model described in ref. [55],
a FRW minisuperspace in N=1
supergravity coupled to locally supersymmetric supermatter (a
massive complex scalar with spin-$\half$ partner) was considered
in ref. [10,11,12].
 In the massless case [11] a ground
 quantum wormhole state can be found
as a solution of the quantum constraints in the form of   an integral
expression.

Locally supersymmetric quantum cosmological models obtained from the more
general N=1 supergravity theory  in the presence of supermatter
were instead considered in ref. [18,21,25,26].
Such  theory is described in detail in ref. [54].
The theory was restricted in ref. [18,21] to the
special case of a $k=1$ FRW
minisuperspace model  with
a family of complex spin-0 scalar fields
 together with their odd (anti-commuting) spin-$\half$ partners.
The Ansatz for the wave function
of the universe was still
found in the ``old'' approach.
 For the two-dimensional
spherically symmetric and flat K\"ahler geometries new
solutions with a simple form for the quantum states were found.
In particular, the
Hartle-Hawking solution is present but  the wormhole state
seems absent\footnote{$^{3}$}{{\sevenrm This issue has been recently
addressed in ref.[62]. The differences in ref. [11,18,21] relatively to the
presence or not of wormhole states can be seen as a consequence of two separate
causes. On the one hand, the choice of Lagrange multipliers (which may simplify
the form of the contraints and corresponding algebra) and on
the other hand how we deal with the fermionic derivative ordering. A suitable
combination of these two aspects seems to make a difference.}}.
Furthermore, different factor
ordering were also considered. In ref. [25] a supermultiplet
constituted by
spin-1 $SU(2)$ gauge fields and their fermionic partners was
added as well to the
supermatter fields. By imposing
the supersymmetry and Lorentz
constraints it was found
that
no physical states were
allowed\footnote{$^{4}$}{{\sevenrm A possible
reason could
either be the use of the ``old'' type of
Ansatz for the wave function of the universe or that the
Ans\"atze for the spin-1 field and
corresponding
supersymmetric partner were not the more general one
under supersymmetry transformations. In fact,  an improved Ansatz for the
Yang-Mills fields seems to allow physical states to be present in a
locally supersymmetric FRW model with spin-1 fields [63].}}.

In this paper we will
study instead a locally supersymmetric
(diagonal) Bianchi type-IX model
coupled to a scalar supermultiplet, formed by
a complex spin-0 scalar fields
together with their odd (anti-commuting)
spin-$\half$ partners. We choose
the corresponding K\"ahler geometry to be a flat one. This
 Bianchi type-IX supersymmetric
model will bear important differences as far as
 FRW models in N=1 supergravity  with
supermatter are concerned.

Firstly, anisotropic gravitational degrees of
freedom are now present. Consequently,
the gravitino fields are no longer required to be
severly restricted to their
spin-$1 \over 2$ modes (see ref. [9-12,18,21,25])
and hence the spin-$3 \over 2$ modes
will play an important role as we will see. In
such a way we hope
our minisuperspace model
with supermatter may be able, in spite of the
inhomogenous modes truncation,  to better
reveal some of the features of the full theory of N=1
supergravity theory with
supermatter.

Secondly, we will take a different approach from the one
involving a direct dimensional
reduction of the $d=4$ theory.
Dimensional reduced one-dimensional models
which inherits invariance under local
time translations, Lorentz and supersymmetry transformations
from   4-dimensional
ones can be obtained
by studying the (more complicated) non-diagonal
Bianchi cases [6,7,15].
To obtain an Ansatz invariant under supersymmetry transfomations one
 must use a non-diagonal triad $e_i^a = b^a_b E_i^b$
where $b_{ab}$ is symmetric ($a,b = 1,2,3$) and combine it with a homogeneous
spatial coordinate
transformation and local Lorentz rotations.
Such an approach was done in ref.
[2-5,6] and
was  recently
 extended for the case of a Bianchi-I model with
a scalar multiplet coupling [26].
However,
it has also been pointed that the dimensional reduction with diagonal Bianchi
models may be safely used
[5,56,57]
 as in the end the algebra of constraints is closed and consistently
a supersymmetric one.
Alternatively,  the differential equations
for our case study
may be obtained instead by studying the quantum
constraints of
the full theory of N=1 supergravity with supermatter
[17,18,54] subject to a
Bianchi type-IX (diagonal) ansatz for the tetrad and correspondingly  to the
gravitino
fields.
Our purposes in applying the
quantum  constraints of the full theory are, on the one hand,
to make a full use of
the recently obtained canonical formulation of N=1
supergravity with supermatter [17],
and on the other hand, to use the (simpler) diagonal
Bianchi-IX metric.
This is a valued approach since the relevant constraints are of first order
in bosonic derivatives, unlike the case in quantized general relativity. We
will
then get expressions for the quantity $\delta \Psi/\delta e_{AA'}^i$ which,
when
evaluated at a Bianchi metric, give the infinitesimal change of the
wave function under a variation of the three scale factors of the Bianchi
type-IX
metric and hence allow the evolution of the wave function subject to the
above mentioned Ansatz to be determined.
The outcome will then
be similar to the one obtained through a dimensional reduction.
We would also like to
stress a particular advantage of using the
(diagonal)
Bianchi IX model as opposed to
more simple cases such as a (general)
Bianchi-I. In the former, we will have
non-zero gravitational potentials
while in the later those will be zero. Hence,
for  the case of  free matter  one  gets
a  model without self interactions for the
Bianchi-I while in the Bianchi-IX case one
expects a behaviour which will be, up to some extent,
more realistic as far as the full theory is concerned.

Finally, we will analyse our Bianchi-IX
supersymmetric model
with supermatter
accordingly to two cases: with and
without a scalar field dependent  analytical potential $P(\Phi)$ in the
Lagrangian.
The fact that we will have all the
homogenous mode components of the gravitino field as well as the terms in
the action corresponding
to
a gravitational potential may allow future
research to address
supersymmetry breaking features from other
points of view. Indeed, supersymmetry breaking phenomena is
related to the behaviour of the scalar field
dependent analytical potential (cf. ref. [54]).
Moreover, the presence of the
analytical potential in the supersymetry
constraints is similar in some sense
to the one induced by a cosmological constant,
therefore allowing for chirality to be broken. It
will be interesting to check in the case $P(\Phi)\neq 0$
if any quantum-physical states are possible.
Nevertheless, one should be very careful as to properly
interpretate the results with respect
to the choice  of wave function Ansatz.

In fact,  let us point out  before proceeding
that we will
follow in this paper an Ansatz for the expansion of the
wave function  in fermionic sectors, which is still
obtained within the ``old'' approach. Basically,
we consider the simplest Lorentz invariants, constructed
only from the irreducible spin-$\half$ and $3 \over 2$ components
of the fermionic fields in presence (see eq.(3.13)). Obviously,
we are aware that this is an overly restricted Ansatz as
compared with the ``new'' approach developed in [23,24]. Our
reasons are as follows.

On the one hand, the so called ``new'' approach conveys
any final calculations of the explicit analytical
solutions of the bosonic amplitudes for the fermionic
middle sectors to a (second order) Wheeler-DeWitt type equation.
It would be much harder to try to solve a Wheeler-DeWitt
type equation for a Bianchi type-IX model
with a complex scalar matter field, which to our knowledge
do not exist yet in the literature (a study of a  Bianchi-IX
model with a real scalar field can be found in [62]).
Moreover, the amplitudes of the
bottom (bosonic) and top (fermionic filled) sectors
in [23,24] are present in
the ``old'' Ansatz procedure, using the
(first order) differential equations  directly obtained
from  the Lorentz and supersymmetry constraints. Furthermore,
the middle sector solution obtained in [23] from the
Wheeler-DeWitt type equation
is not new in the sense that it was  allready
found out using a specific  definition of
homogeneity conditions for the gravitinos [8]
and  in the context of the
``old'' Ansatz construction.

On the other hand, we would like to obtain new solutions
which would bear any physical significance
with regard to  minisuperspace Bianchi-IX cosmologies.
A straightforward
(although tedious) way would be to consider the
(first order) set of differential equations obtained
from the supersymmetric constraints, as the ``old''
procedure allows. We hope in this way to get solutions that
ought to  be present as well in the approach of [23,24], possibly
giving us an insight on how to generalize it to
couplings with supermatter. As we will see, solutions
in the top and bottom sectors as well as in
some middle sectors are obtained for our model when
the analytical potential is zero. We suspect that these or similar
properties and solutions would be obtained in the ``new'' approach
but at the expense of having to
deal with a Wheeler-DeWitt equation.
Of course, we could possibly find out  other solutions
with a more
general Lorentz invariant Ansatz for the wave function of the
universe. Our point is that the ``old'' Ansatz construction
may still be usefull, up  to a certain point, if
addressed within its proper context.

The above statements should however be taken with care as far as
the inclusion of chirality  breaking  terms (like a cosmological constant)
are concerned. The non-existence of physical states in that
case is again another indication  that the ``old'' restrictive
Ansatz is not adequate when dealing with these type of situations.
Surely the ``new'' approach is a welcomed feature for the
theory of canonical quantization of
locally supersymmetric models.

This paper is then organized as follows.
In section II we briefly describe the canonical
formulation of N=1 supergravity with supermatter,
following ref. [17,18]. For the sake of
completeness we present the general theory with gauge
supermatter, ie, with a family of spin-0 and 1
fields and their fermionic partners.
Some basic features are improved as well. More precisely,
the variation of some physical observables under
supersymmetry transformations obtained
from the supersymmetry constraints generators in [17,18]
do {\it not} coincide
with the ones present in the
literature (see, e.g. ref. [54]) for the cases
of the spin-$1 \over 2$ partner
of the scalar field and the gravitino.
We explain why this problem was present and
suggest the inclusion of
adequate terms in the
Hamiltonian formulation as to get consistent results with the
usual supersymmetry
transformations.  The quantization of the diagonal Bianchi-IX
model is studied in section III,
by studying the quantum constraints in the full theory
subject to a suitable Bianchi-IX Ans\"atze for the
fields. The chosen supermatter model will be restricted
to a spin-0 and spin-$1 \over 2$ supermultiplet,
corresponding to a 2-dimensional flat K\"ahler geometery.
We will discuss separatly two cases:
 when the  scalar field dependent
analytic potential $P(\Phi)$ is arbitrarly and
  when  $P(\Phi)$ is zero.
For our Lorentz invariant Ansatz of
the wave function of the universe
 we will
find that in the later the wave function  has a simple form. Namely, the only
non zero
components of the wave function can be found in the sectors
with no fermions (bosonic) and in three other sectors, more precisely filled
with just the spin-$1 \over 2$
fermionic partners of the scalar field, another  filled with just the spin-$1
\over 2$ and
$3 \over 2$ mode components of the spatially homogeneous gravitino field
and finally one totally filled with spin-$1 \over 2$
fermionic partners of the scalar field as well as the
the spin-$1 \over 2$ and
$3 \over 2$ mode components of the gravitino field.
In section IV we present our discussons and conclusions.

 \bigskip

 \bigskip

\bigskip

\medskip

{\bf II. Canonical formulation of N=1 supergravity with supermatter}

The Lagrangian of the more general theory of N=1 supergravity with supermatter
 is given in eq.~(25.12) of [54] and  it is too long to write out
here. It depends on the tetrad
$ e^{A A'}_{~~~~\mu} $, where $ A, A' $ are two-component
indices
using the conventions of [1]
and $ \mu $ is a space-time index, the  spin-${3 \over 2} $
gravitino field $
\( \psi^A_{~~\mu}, \wti \psi^{A'}_{~~\mu} \) $.
whose
components
are taken to be odd elements of a Grassmann algebra,
on a vector field $ A^{(a)}_\mu $
labelled by a group  index $ (a) $, its  spin-$\half $ partner $ \(
\lam^{(a)}_A, \wti \lam^{(a)}_{A'} \) $, a family of scalars $ \( \Phi^I,
\Phi^{J^*}
\) $ and their  spin-$\half$ partner $ \( \rchi^I_A, \wti \rchi^{J
^*}_{A'} \) $. The indices $ I, \ldots, J^*, \ldots $ are K\"ahler indices, and
there is a K\"ahler metric
$$ \rg_{I J^*} =   K_{I J^*} \eqno (2.1) $$
on the space of $ \( \Phi^I, \Phi^{J^*} \) $, where $ K_{I J^*} $ is a
shorthand
for $ \pt^2 K / \pt \Phi^I \pt \Phi^{J^*} $ with $ K $ the K\"ahler potential.
Each
index $ (a) $ corresponds to an independent Killing vector field of the
K\"ahler geometry. Such Killing vectors are holomorphic vector fields:
$$ \eqalignno {
X^{(b)} &= X^{I (b)} \( \Phi^J \)~{\pt \over \pt \Phi^I}~,&(2.2a) \cr
X^{^* (b)} &= X^{I^* (b)} \( \Phi^{J^*} \)~{\pt \over \pt \Phi^{I^* }}~.
&(2.2b) \cr
} $$
Killing's equation implies that there exist real scalar functions $ D^{(a)}
\( \Phi^I, \Phi^{I^*} \) $ known as Killing potentials, such that
$$ \eqalignno {
\rg_{I J^*} X^{J^* (a)} &= i~{\pt \over \pt \Phi^I}~D^{(a)}~,&(2.3a) \cr
\rg_{I J^*} X^{I (a)} &= - i~{\pt \over \pt \Phi^{J^*}}~D^{(a)}~.
&(2.3b) \cr
} $$

In the Hamiltonian decomposition, the variables are split into
 $ e^{A A'}_{~~~~i},~\psi^A_{~~i},~\wti \psi^{A'}_{~~i},~A^{(a)}_i,~$
($i=1,2,3$, denotes  spatial
components)
$\lam^{(a)}_A,~\wti \lam^{(a)}_{A'},~\rchi^I_A,~\wti
\rchi^{J^*}_{A'},~\Phi^I,~\Phi^{J^*} $, which together with the bosonic momenta
are the basic dynamical variables of the theory, and the
Lagrange multipliers $ N,~N^i,~\psi^A_{~~0},~\wti
\psi^{A'}_{~~0},~A^{(a)}_0,~M_{A B},~\wti M_{A' B'} $, where $ N,
N^i $ are formed from the $ e^{A A'}_{~~~~0} $ and the $ e^{A A'}_{~~~~i} $
[1], and $ M_{A B}, \wti M_{A' B'} $  involve the zero components $ \om_{A B
0}, \wti \om _{A' B' 0} $ of the connection (see below).
The constraint generators are functions of the
basic dynamical variables.

The total  Hamiltonian  has the form
$$ \eqalignno {
H = N \cH_\perp + N^i \cH_i + \psi^A_{~~0} S_A &+ \wti S_{A'} \wti
\psi^{A'}_{~~0} \cr
+ A^{(a)}_0 Q_{(a)} + M_{A B} J^{A B} &+ \wti M_{A' B'} \wti J^{A' B'}~, &(2.4)
\cr } $$
expected for a theory with the corresponding gauge invariances. Here $ N $
and $ N^i $ are the lapse function and shift vector, while $ \cH_\perp $
and $ \cH_i $ are the (modified) generators of deformations in the normal and
tangential directions. $ S_A $ and $ \wti S_{A'} $ are the local supersymmetry
generators, $ Q_{(a)} $ is the generator of gauge invariance, and $ J^{A B} $
and $ \wti J^{A' B'} $ are the generators of local
Lorentz rotations.
Classically, all the
constraints $ \cH_\perp, \cH_i, S_A, \wti S_{A'}, Q_{(a)}, J^{A B}, J^{A' B'}$
{}~vanish, and form a  set of (first-class)
constraints,  satisfying  an algebra.
Quantum mechanically, the constraints become
operators which annihilate physical states $ \Psi $:
$$ \eqalignno {
\cH_\perp \Psi &= 0~, \ \ \ \ \ \ \cH_i \Psi = 0~, \ \ \ \ \ \ S_A \Psi = 0~,
\ \ \ \ \ \ \ol S_{A'} \Psi = 0~, \cr
Q_{(a)} \Psi &= 0~, \ \ \ \ \ \ J^{A B} \Psi = 0~, \ \ \ \ \ \ \ol J^{A' B'}
\Psi =0~. &(2.5) \cr
} $$

For the
gravitino and spin-$\half$ fields, the canonical momenta give second-class
constraints of the types described in [1,43,58]. These are eliminated when
Dirac
brackets are introduced [1,10,43] instead of the original Poisson brackets. In
particular, one obtains nontrivial Dirac brackets for $ p_{A A'}^{~~~~i} $,
(the momentum conjugate to $ e^{A A'}_{~~~~i} $), for $ \psi^A_{~~i} $ and $
\wti \psi^{A'}_{~~i} $, for $ \lam^{(a)}_A $ and $ \wti \lam^{(a)}_{A'} $, for
$ \rchi^I_A $ and $ \wti \rchi^{J^*}_{A'} $, and for $ \pi_L,~\pi_{L^*} $,
(the momenta conjugate to $ \Phi^L,~\Phi^{L^*} $). These can be made into
simple
brackets as follows.

The brackets involving $ p_{A A'}^{~~~~i},~\psi^A_{~~i} $ and $ \wti
\psi^{A'}_{~~i} $ can be simplified as in the case of pure $ N = 1 $
supergravity [1] by  redefining
$$ p_{A A'}^{~~~~i} \to \hat p_{A A'}^{~~~~i} =
p_{A A'}^{~~~~i} -
{1 \over
\sqrt 2} \eps^{i j k} \psi_{A j} \wti \psi_{A' k}~. \eqno (2.6) $$
The $ \Phi^K $ and $ \Phi^{K^*} $ dependence of $ K_{I J^*} $ is responsible
for
 unwanted Dirac brackets among $ \rchi^I_A,~\wti \rchi^{J^*}_{A},~\pi_L $
and $ \pi_{L^*} $. In fact,  defining $ \pi_{I A} $ and
$ \wti \pi_{I^* A'} $
to be the momenta conjugate
to $ \rchi^{I A} $, and $ \wti \rchi^{I^* A'} $, respectively, one has
$$ \eqalignno {
\pi_{I A} + {i h^{{1 \over 2}}
\over \sqrt 2}~K_{I J^*} n_{A A'} \wti \rchi^{J^* A'} &= 0~,
\cr
\wti \pi_{J^* A'} + {i h^{{1 \over 2}} \over \sqrt 2}~K_{I J^*} n_{A A'}
\rchi^{I A} &= 0~.
&(2.7) \cr } $$
Here $ n^{A A'} $ is the spinor version of the unit
future-directed normal vector $ n^\mu $, obeying
$$ n_{A A'} n^{A A'} = 1~, \ \ \ \ \ \ n_{A A'} e^{A A'}_{~~~~i} = 0~. \eqno
(2.8) $$
One cures this by introducing  the modified variables
$$ \eqalignno {
\hat \rchi_{I A} &= h^{1 \over 4} K^{1 \over 2}_{I J^*} \de^{K J^*} \rchi_{K
A}~, \cr
\hat {\wti \rchi}_{I^* A'} &= h^{1 \over 4} K^{1 \over 2}_{J I^*} \de^{J
K^*} \wti \rchi_{K^* A'}~, &(2.9) \cr
} $$
where the factor of $ h^{1 / 4} $ has been introduced for later use
(in the
time gauge -- see below).
$K^{1 \over 2}_{I J^*} $ denotes a  ``square root'' of the K\"ahler
metric, obeying
$$ K^{1 \over 2}_{I J^*} \de^{K J^*} K^{1 \over 2}_{K L^*} = K_{I L^*}~. \eqno
(2.10) $$
This may be found by diagonalizing $ K_{I J^*} $ via a unitary transformation,
assuming that the eigenvalues are all positive. One needs to assume that
there is an ``identity metric'' $ \de^{K J^*} $ defined over the K\"ahler
manifold; this will be true if a positive-definite vielbein field can be
introduced. Then the second-class constraints in Eq.~(2.7) read
$$ \eqalignno {
\hat \pi_{I A} &+ {i \over \sqrt 2}~\de_{I J^*}~n_{A A'} \hat {\wti
\rchi}^{J^* A'} = 0~, \cr
\hat {\wti \pi}_{I^* A'} &+ {i \over \sqrt 2}~\de_{I J^*}~n_{A A'} \hat
\rchi^{J A} = 0~. &(2.11) \cr } $$
Finally,  the brackets among $ \hat p_{A
A'}^{~~~~i},~\lam^{(a)}_A,~\wti \lam^{(a)}_{A'},~\hat \rchi^I_A $ and $
\hat {\wti \rchi}^{J^*}_{A'} $ are dealt with by defining
(see ref. [58,59])
$$ \hat \lam^{(a)}_A = h^{1 \over 4} \lam^{(a)}_A~, \ \ \ \ \hat {\wti
\lam}^{(a)}_{A'} = h^{1 \over 4} \wti \lam^{(a)}_{A'}~, \eqno (2.12) $$
and
then going to the time gauge. In this case,
 the tetrad component $ n^a $
of the normal vector $ n^\mu $ is henceforward restricted by
$$ n^a = \de^a_0~, \eqno (2.13) $$
or equivalently
$$ e^0_{~~i} = 0~. \eqno (2.14) $$
Thus the original Lorentz rotation freedom becomes replaced by that of
spatial rotations. In the time gauge, the geometry is described by the triad
$ e^\al_{~~i} (\al = 1, 2, 3) $, and the conjugate
momentum\footnote{$^{3}$}{Notice that
$   \pi^{ij} \equiv - \half p^{(ij)} = \half e^{AA'(i}p_{AA'}^{j)}
=  - \half e^{\alpha(i}p_{\alpha}^{j)}$
where the last equality
follows from the time gauge conditions; see ref. [43,58].}
is $
p_\al^{~~i} $.

The resulting Dirac brackets are
$$ \eqalignno {
\[  p_\al^{~~i},~\psi^B_{~~j} \]_D &= 0~, \cr
\[  p_\al^{~~i},~\wti \psi^{B'}_{~~j} \]_D &= 0~, \cr
\[  p_\al^{~~i}, \ \hat p_\beta^{~~j} \]_D &= 0~, \cr
\[  p_\al^{~~i}, \ \hat \lam^{(a)}_A \]_D &= 0~, \cr
\[  p_\al^{~~i}, \ \hat \rchi^I_{~~A} \]_D &= 0~, \ {\rm etc.}
 &(2.15) \cr
\[ \pi_L, \ \pi_M \]_D &= 0~, \cr
\[ \pi_L, \ \hat \rchi^A_{~~I} \]_D &= 0~, \ {\rm etc.}
&(2.16)
\cr } $$
The remaining brackets are standard; the nonzero fermionic brackets are
$$ \eqalignno {
\[ \hat \lam^{(a)}~_A(x), \ \hat {\wti \lam}^{(b)}_{~~A'}(x) \]_D &=
\sqrt 2 i n_{A A'} \de^{(a) (b)} \de \( x, x' \)~, &(2.17) \cr
\[ \hat \rchi^I_{~~A} (x), \ \hat {\wti \rchi}^{J^*}_{~~A'} \( x' \) \]_D
&= \sqrt 2 i n_{A A'} \de^{I J^*} \de \( x, x' \)~, &(2.18) \cr
\[ \psi^A_{~~i} (x), \ \wti \psi^{A'}_{~~j} \( x' \) \]_D &= {1 \over
\sqrt 2} D^{A A'}_{~~~~i j} \de \( x, x' \)~, &(2.19) \cr } $$
where
$$ D^{A A'}_{~~~~i j} = - 2 i h^{- \half} e^{A B'}_{~~~~j} e_{B B' i} n^{B
A'}~.
\eqno (2.20) $$
We would like to point out that some of the signs and factors
in the above expressions are different from the ones in
ref. [1]. The quantum representation  of momenta and coordinate will
be done accordingly, in order to achieve self-consistency
(see section III).

Let us now address the supersymmetry constraints from the
Hamiltonian formalism. We will follow closely the framework
presented in ref.[1,43] and adequality extended it to the case
where gauged supermatter is present. As stated above,
the momenta conjugate to the basic
fields are found from the Lagrangian  and
leading to a number of primary and secondary constraints, these last
characteristic of systems with fermions. Eliminating these
and obtaining modified bracket relations, as explained, we
obtain an explicit form for the terms present in
 expression (2.4). However, the full Hamiltonian
(2.4) contains arbitrary Lorentz rotations and these ought
to be included through  a geometrically meaningfull Lagrangian
multipliers, namely the time component $\omega_{AB0},
\tilde{\omega}_{A'B'0}$ of the connection forms multiplied
by a minus sign [1,43]. Choosing to implement this procedure,
the improved constraints in (2.4) will then differ from the previous
ones by terms proportional to projections of
$J_{AB}$ or its derivatives. The parts in $J_{AB},\wti J_{A'B'}$
corresponding to the spin$-2$ and $3 \over 2$ fields will contribute
to $S_A,\bar S_{A'}$ as in [1,43], after expanding the
spatial covariant derivative in its torsion-free and contorsion
parts.
The inclusion of components from $J_{AB}$
which depend on the mater fields will contribute
with new terms of the type $\psi\chi\bar\chi,~~ \psi\lambda
\bar\lambda$ and their hermitian conjugates to the supersymmetry
constraints [43]. We notice that this last step is missing in the
procedure employed in ref. [17]. In the end of this section we further discuss
the implications of its absence and which problems its presence might solve.

The supersymmetry constraint $ \wti S_{A'} $ is then found to be
$$ \eqalign {
\wti S_{A'} =
&
- \sqrt 2 i   e_{A A' i} \psi^A_{~~j}  \pi^{i j}
+
\sqrt 2
\eps^{i j k} e_{A A' i}~^{3 s} \wti D_j \psi^A_{~~k}
\cr
} $$
$$
+ {1 \over \sqrt 2}
\left [
\eqalign {
&
\pi_{J^*}
+
{i \over \sqrt 2} h^{{1 \over 2}} g_{L M^*} \Ga^{M^*}_{J^* N^*} n^{B B'} \wti
\rchi^{N^*}_{~~~B'} \rchi^L_{~~B}
\cr
-
&
 {i h^{{1 \over 2}} \over 2 \sqrt 2} K_{J^*} g_{M M^*} n^{B B'} \wti
\rchi^{M^*}_{~~~B'}
\rchi^M_{~~B}
-
{1 \over 2 \sqrt 2} \eps^{i j k} K_{J^*} e^{B B'}_{~~~~j} \psi_{k B} \wti
\psi_{i B'}
\cr
- &
w_{[1]}\sqrt 2 h^{{1 \over 2}} g_{I J^*} \rchi^{I B} e_{B B'}^{~~~~m} n^{C B'}
\psi_{m C} \cr }
\right ]
\wti \rchi^{J^*}_{~~~A'}
$$
$$ \eqalignno {
-
&
\sqrt 2 h^{{1 \over 2}} g_{I J^*} \( \wti \cD_i \Phi^I \) \wti \rchi^{J^*}_{B'}
n^{B B'}
e_{B A'}^{~~~~i} +
w_{[2]}{i \over 2} g_{I J^*} \eps^{i j k} e_{A A' j} \psi^A_{~~i} \wti
\rchi^{J^*
B'} e_{B B' k} \rchi^{I B} \cr
+ &w_{[3]}{1 \over 4} h^{{1 \over 2}}
\psi_{A i} \( e_{B A'}^{~~~~i} n^{A C'} - e^{A C' i} n_{B A'}
\) g_{I J^*} \wti \rchi^{J^*}_{C'} \rchi^{I B}
\cr
-
&
h^{{1 \over 2}} \exp ( K / 2 ) \[ 2 P n^A_{~~A'} e_{A B'}^{~~~~i} \wti
\psi^{B'}_{~~i} + i \( D_I P \) n_{A A'} \rchi^{I A}
\] \cr
&
- {i \over \sqrt 2} \pi^{n (a)} e_{B A' n} \lam^{(a) B}
+ {1 \over 2 \sqrt 2} \eps^{i j  k} e_{B A' k} \lam^{(a) B} F^{(a)}_{i j}
+ {1 \over \sqrt 2} h^{{ 1 \over 2}} g D^{(a)} n^A_{~~A'} \lam^{(a)}_{~~~A}
\cr
+
&
w_{[4]}{1 \over 4} h^{{1 \over 2}} \psi_{A i} \( e_{B A'}^{~~~~i} n^{A C'} -
e^{A C' i} n_{B
A'} \) \wti \lam^{(a)}_{C'} \lam^{(a) B} \cr
-
&
{i  \over 4} h^{{1 \over 2}} n^{B B'} \lam^{(a)}_{~~~B} \wti \lam^{(a)}_{~~~B'}
K_{J^*} \wti \rchi^{J^*}_{~~~A'} ,
&(2.21)
\cr
} $$
where $ \lam^{(a)}_A,~\wti \lam^{(a)}_{A'} $ and $ \rchi_{I A},~\wti
\rchi_{I^* A'} $ should be redefined as in Eqs.~(2.9), (2.12).
The other supersymmetry constraint ($S_A$) is just the
hermitian conjugate of (2.21).
Here $ e_{A A' i} = \sig^\al_{A A'} e_{\al i} $ [1], where $ \sig^\al_{A A'}
(\al
= 1, 2, 3) $ are Infeld-van der Waerden symbols and $  \pi^{i j} = -
{1 \over 2} e^{\al ( i} \hat p_\al^{~~j )} $. Notice that the
three last lines in (2.21) correspond to the presence of a spin-1 field and
fermionic partner in the supermatter content, i.e., if gauged
supermatter is considered [54,60]. These terms will
{\bf not} be used in our Bianchi type-IX model.
The  $w_{[i]}, i=1,2,3,4$ denote numerical coefficients
which correspond to the inclusion of the terms
 $\psi\chi\bar\chi,~~ \psi\lambda
\bar\lambda$ and their hermitian conjugates to the supersymmetry
constraints via  $\omega_{AB0}J^{AB}$ and hermitian
conjugate.
Also [54]
$$ \eqalignno {
^{3 s} \wti \cD_j \psi^A_{~~k} =
&\pt_j \psi^A_{~~k} + ~^{3 s} \om^A_{~~B j}
\psi^B_{~~k} \cr
+ &{1 \over 4} \( K_K \wti \cD_j \Phi^K - K_{K^*} \wti \cD_j \Phi^{K^*} \)
\psi^A_{~~k} \cr
+ &{1 \over 2} g A^{(a)}_j \( I m F^{(a)} \) \psi^A_{~~k}~, &(2.22) \cr } $$
where $ ^{3 s} \om_{A B j}, \ ^{3 s} \wti \om_{A' B' j} $ give the
torsion-free three-dimensional connection, and
$$ \wti \cD_i A^K = \pt_i A^K - g A^{(a)}_i X^{K (a)}~, \eqno (2.23) $$
with $ g $ the gauge coupling constant and $ X^{K (a)} $ the $ a $th Killing
vector field. Furthermore, the analytic functions $ F^{(a)} \( \Phi^J
\) $ and $ F^{^* (a)} \( \Phi^{I^*} \) $ arise [54,60]
from the transformation of the
K\"ahler potential $ K $ under an isometry generated by the Killing vectors $
X^{(a)} $ and $ X^{^*(a)} $. Also, in Eq.~(2.21),
$\pi^{n (a)} $ is the momentum conjugate to $ A^{(a)}_n ,~
K_{J^*} $ denotes
$ \pt K / \pt \Phi^{J^*},~\Ga^{M^*}_{J^* N^*} $ denotes the
Christoffel
symbols [54,60] of the K\"ahler geometry, and $ P = P \( \Phi^I \) $
gives the anlytical
potential of the theory.

The gauge generator $ Q^{(a)} $ is given classically by
$$ \eqalignno {
Q^{(a)} &= - \pt_n \pi^{n (a)} - g f^{a b c} \pi^{n (b)} A^{(c)}_n \cr
&+ g \( \pi_I X^{I (a)} + \pi_{I^*} X^{I^* (a)} \) \cr
&+ \sqrt 2 i h^{{1 \over 2}} g K_{M I^*} n^{A A'} X^{J^* (a)} \Ga^{I^*}_{J^*
N^*} \wti
\rchi^{N^*}_{A'} \rchi^M_A \cr
&- \sqrt 2 i h^{{1 \over 2}} g n^{A A'} \wti \lam^{(b)}_{A'} \[ f^{a b c}
\lam^{(c)}_A +
{1 \over 2} i \( I m F^{(a)} \) \lam^{(b)}_A \] \cr
&+ \sqrt 2 i h^{{1 \over 2}} g n^{A A'} K_{I J^*} \wti \rchi^{J^*}_{A'}
\[ {\pt X^{I (a)}
\over \pt \Phi^J} \rchi^J_A + {1 \over 2} i \( I m F^{(a)} \) \rchi^I_A \] \cr
&- {i \over \sqrt 2} g \( I m F^{(a)} \) \eps^{i j k} \wti \psi_{i A'} e^{A
A'}_{~~~~j} \psi_{A k}~, &(2.24) \cr } $$
where $ f^{a b c} $ are the structure constants of the isometry group.

It is worthwhile to notice that we expect now to
obtain the correct transformation properties (cf. ref. [54])
of the
physical fields under both
supersymmetry transformations,
using  brackets $\delta_{\xi}\psi^A_i
\equiv [\wti \xi_{A'} \wti S^{A'}, \psi^A_i]_D$, etc, where
$\xi$ is a constant spinor parametrizing the
supersymmetric transformation.
In fact, that
 was not possible for some fields, when using  the explicit form
for the supersymmetry constraints in ref. [17,18] as it can
be checked. The reasons
are as   follows. On the one hand, the matter terms in the
Lorentz constraints $J_{AB}, \wti
J_{A'B'}$ were not included in the supersymmetry
constraints, following and extending the framework
presented in [43]. On the other hand, expressions
 only valid in pure N=1 supergravity
were employed to simplify the supersymmetry constraints with
supermatter. Namely, the expressions for $S_A = 0,
\wti S_{A'} = 0$ in pure N=1 supergravity
were used to re-write the spatial covariant
derivative $^3 D_i$ in terms of its torsion free part $^{3s} D_i$and
remaining terms which include the contorsion. When supermatter is present, we
expect
the different matter fields to play a role in the Lorentz constraints terms
which ought to
be included in the supersymmetry
constraints once $\omega_{AB}^0 J^{AB}$ and its hermitian conjugate are
employed in the
canonical action. A similar argument applies when we expand $^3 D_i$ but using
(2.22), (2.23).

\bigskip

\medskip

{\bf III. Quantization of the diagonal Bianchi type-IX model}

In this section we study the  Bianchi type-IX model   with
spatial metric in diagonal form, using the supersymmetry constraints
derived in the previous section. We restrict our case study
to a supermatter model constituted only by a scalar field and
its spin$-{1 \over 2}$ partner with a two-dimensional
flat K\"ahler geometry.
The K\"ahler potential  would be just  $\phi \bar \phi$,
the K\"ahler metric is $g_{\phi \bar \phi} = 1$ and
the Levi-Civita connections are zero.
The scalar super-multiplet, consisting of a complex massive scalar
field $ \phi $ and massive spin-$\half$ field $ \rchi_A, \bar\rchi_{A'}$
are chosen to be
spatially homogeneous, depending only on time.

The 4-metric $ g_{\mu \nu} $
of diagonal Bianchi type-IX
is given by
$$ g_{\mu \nu} = \eta_{a b} e^a_{~\mu} e^b_{~\nu}~, \eqno (3.1) $$
where $ \eta_{a b} $ is the Minkowski metric $ (\mu, \nu = 0, \ldots, 3 ; a,
b = 0, \ldots, 3) $ and the non-zero components of the tetrad $ e^a_{~\mu} $
are given by
$$ \eqalign {
e^0_{~0} = N~,~~e^1_{~0} = a_1 N^i E^1_{~i}~,~~e^2_{~0} = &a_2 N^i E^2_{~i}~,~~
e^3_{~0} = a_3 N^i E^3_{~i}~, \cr
e^1_{~i} = a_1 E^1_{~i}~,~~~~~~e^2_{~i} = &a_2 E^2_{~i}~,
{}~~~~~~e^3_{~i} = a_3
E^3_{~i}~. \cr }
\eqno (3.2) $$
Here $ E^1_{~i}, E^2_{~i}, E^3_{~3} (i = 1, 2, 3) $ are
a basis of unit
left-invariant one-forms on the three-sphere
[44] and $ N, N^i, a_1, a_2, a_3 $ are spatially constant.
We can also write
$$ h_{i j} = a_1^2 E^1_{~i} E^1_{~j} + a_2^2 E^2_{~i} E^2_{~j} +
a_3^2 E^3_{~i}
E^3_{~j}~. \eqno (3.3) $$
In the calculation, we shall repeatedly
need the expression:
$$ \eqalignno {
\om_{A B i} n^A_{~~B'} e^{B B' j} &= {i \over 4} \( {a_3 \over a_1a_2}
+ {a_2 \over a_3
a_1} - {a_1 \over a_2 a_3} \) E^1_{~i} E^{1 j} \cr
{}~&+ {i \over 4} \( {a_1 \over a_2 a_3} +
{a_3 \over a_2 a_1} - {a_2 \over a_3 a_1} \) E^2_{~i}
E^{2 j} \cr
{}~&+ {i \over 4} \(
{a_2 \over a_3 a_1} + {a_1 \over a_2 a_3} - {a_3 \over a_1 a_2} \) E^3_{~i}
E^{3 j}. &(3.4) \cr } $$
We require that the
components
$ \( \psi^A_{~~0}, \ol \psi^{A'}_{~~0} \) $ be functions of time only. We
further require that $ \psi^A_{~~i} $ and $ \ti \psi^{A'}_{~~i} $ be
spatially homogeneous in the basis $ e^a_{~i} $.

We now proceed to solve the supersymetry and Lorentz
constraints for the case of a diagonal
Bianchi-IX model with a scalar supermultiplet.
The $ \cH_\perp $ and $ \cH_i $ constraints can be defined
through the anti-commutator of $ S_A $ and $ \ol S_{A'} $, as in the case of $
N = 1 $ supergravity without matter fields [1,22,23,24].
Thus the remaining
constraints imply $ \cH_\perp \Psi = 0,~\cH_i \Psi = 0 $; if one could find a
solution of the remaining quantum constraints, the $ \cH_\perp $ and $ \cH_i
$ constraints would follow (with a certain choice of factor-ordering).

A quantum description can be made by studying (for example)
Grassmann-algebra-valued wave functions of the form
$ \Psi \[ e^{AA'}_{~~i}, \ \psi^A_{~~i},
\bar \rchi_A,
\ \phi,
\ \ol\phi \] $.
The choice of $ \bar \rchi_A \equiv n_A^{~~A'} {\bar \rchi}_{A'} $
rather than $
\rchi_A $ is designed so that the quantum constraint $ \ol S_{A'} $ should
be of first order in momenta (cf. also ref.
[9-12,18,21,25]).
For simplicity, we have droped the hat ``$\land$'' henceforth and
apply our homogeneous Bianchi-IX Ansatze consistently
throughout the paper.

The momenta are represented by
$$ \eqalignno {
p_{A A'}^{~~~~i} &\to - i \hbar {\de \over \de e^{A A'}_{~~~~i}}
- {1 \over \sqrt{2}}
\eps^{i j k} \psi_{A j} \bar \psi_{A' k}~,&(3.5) \cr
\pi_\phi &\to - i \hbar {\pt \over \pt \phi}~, &(3.6) \cr
\pi_{\ol\phi} &\to - i \hbar {\pt \over \pt \ol \phi}~, &(3.7) \cr
\ol \psi^{A'}_{~~i} &\to  {1 \over \sqrt 2} i \hbar D^{A A'}_{~~~~j i}
 h^{1 \over 2}
{\pt \over \pt
\psi^A_{~~j}}~, &(3.8) \cr
\rchi^{A} &\to - \sqrt 2 \hbar
 {\pt \over \pt
 \ol \rchi^A}~.
&(3.9) \cr } $$
We have made
the replacements $ \de \Psi / \de \psi^B_{~~j} \longrightarrow h^{1 \over 2}
\pt \Psi / \pt \psi^B_{~~j} $,
$\delta \Psi / \delta \ol \rchi^A \to \pt \Psi /\pt  \ol \rchi^A$,
where $ \pt / \pt \psi^A_{~~j}, \pt /
\pt \ol \rchi^A$ denotes left differentiation.
It is worthwhile  to make the following
comments about these   replacements.
The $ h^{1 \over 2} $ factor is necessary as to ensure that each
term has the correct weight in the equations, namely when one takes a
variation of a  Bianchi geometry whose spatial sections are compact,
multiplying by
 $ \de  / \de h_{i j} $ and integrating over the
three-geometry [6,7,14,15,16]: the cause can be
identified in the term $h^{-{1\over 2}}$ in expression (2.20).
One can  check, e.g., that the inclusion of $h^{{1 \over 2}}$
gives the correct supersymmetry constraints in the $ k = + 1 $
Friedmann model, where the model was quantized using
the alternative approach via a supersymmetric Ansatz
[9-12,21,25].
It is interesting as well to notice that for the $\rchi,\ol \rchi$
fields no such requirements seemed to be needed
to establish the
equations for the bosonic amplitudes of the
wave function of the universe.

The supersymmetry constraints become then, in differential
operator form
$$ \eqalignno{   \ol S_{A'} &=
 -i\sqrt{2} \[-i \hbar {1\over 4}\[e_{AA'i} \psi^A_j\] e^{CC'i}
{\de \over \de e^{CC'j}}
 -i \hbar {1\over 4}  \[e_{AA'i} \psi^A_j\]   e^{CC'j}
{\de \over \de e^{CC'i}}
 \]
\cr
&
+\sqrt{2} \epsilon^{ijk} e_{AA'i} ~^{3(s)}\om^A_{~~Bj} \psi^B_{~~j}
\cr
&
-{i \over \sqrt{2}}\hbar n_{CA'} \ol \rchi^C {\pt \over \pt \ol \phi}
\cr
&
- i \sqrt{2} \hbar h e^{K/2} P(\phi) n^A_{~~A'} e_{~AB'}^i D^{CB'}_{~~~ji}
{\pt \over \pt \psi^C_j}
\cr
&
- i \sqrt{2} \hbar h e^{K/2} D_\phi P ~ n^A_{~~A'}
{\pt \over \pt \ol \rchi^A}
\cr
&
-{i \over 2\sqrt{2}} h^{{1 \over 2}} \hbar \phi n^{BB'} n_{CB'}
\ol \rchi^C n_{DA'} \ol \rchi^D { \pt \over \pt \ol \rchi^B}
\cr
&
-{i \over 4\sqrt{2}} \hbar h^{{1 \over 2}} \phi \epsilon^{ijk}
e^{BB'}_{~~~j} \psi_{kB} D^C_{~~B'li} n_{DA'}\ol\rchi^D
{\pt \over \pt \psi^C_l}
\cr
&
- \hbar \sqrt{2} h^{{1 \over 2}}
e^{B~~~m}_{~~B'} n^{CB'} \psi_{mC} n_{DA'} \ol \rchi^D
{ \pt \over \pt \ol \rchi^B}
\cr
&
- {i \over \sqrt{2}} \hbar \epsilon^{ijk} e_{AA'j} \psi^A_i
n_{DB'} \ol\rchi^D e^{BB'}_{~~~k}  { \pt \over \pt \ol \rchi^B}
\cr
&
+ {1 \over 2\sqrt{2}} \hbar h^{{1 \over 2}} \psi_{iA}
(e^{B~~~i}_{~~A'} n^{AC'} - e^{AC'i} n^B_{~~A'})
n_{DC'} \ol \rchi^D { \pt \over \pt \ol \rchi^B},
& (3.10)\cr}
$$

$$ \eqalignno{S_{A} &=
 i\sqrt{2} \[
 - \hbar^2 {1\over 4 \sqrt{2}}
\[e_{AA'i} h^{{1 \over 2}} D^{PA'}_{~~~kj}
{\pt \over \pt \psi^P_{~~k}}\]
 e^{CC'i}
{\de \over \de e^{CC'j}}
\right. \cr
& \left.  - \hbar^2 {1\over 4 \sqrt{2}}
\[e_{AA'i} h^{{1 \over 2}} D^{PA'}_{~~~kj}
{\pt \over \pt \psi^P_{~~k}}\]
e^{CC'j}
{\de \over \de e^{CC'i}} \]
\cr
&
- i\hbar \epsilon^{ijk} h^{{1 \over 2}}  e_{AA'i} ~^{3(s)}\ol\om^{A'}_{~~B'j}
D^{PB'}_{~~~mk} {\pt \over \pt \psi^P_{~m}}
\cr
&
+   {i }\hbar^2  {\pt \over \pt \ol \rchi^A} {\pt \over \pt  \phi}
\cr
&
-  h^{{1 \over 2}} e^{K/2} P(\phi) n_A^{~~A'} e_{BA'}^{~~~i} \psi^B_{~~i}
\cr
&
+ i  h^{{1 \over 2}} e^{K/2} (D_\phi P)^* ~ n_A^{~~A'} n_{CA'}\ol \rchi^C
\cr
&
-{i \over 2} h^{{1 \over 2}} \hbar^2 \ol\phi n^{BB'} n_{DB'}
\ol \rchi^D {\pt \over \pt  \ol \rchi^A} { \pt \over \pt \ol \rchi^B}
\cr
&
+ {i \over 4} \hbar^2 h^{{1 \over 2}} \ol\phi \epsilon^{ijk}
e^{BB'}_{~~~j} \psi_{iB} D^C_{~~B'mk} {\pt \over \pt \ol\rchi^A}
{\pt \over \pt \psi^C_m}
\cr
&
+ i \hbar^2  h
e_{B}^{~~B'm} n^{BC'} D^C_{~~C'km}  n_{DB'} \ol \rchi^D
{ \pt \over \pt \ol \rchi^A} {\pt \over \pt \psi^C_k}
\cr
&
+ {i \over 2} \hbar^2 \epsilon^{ijk} h^{{1 \over 2}}
e_{AA'j} D^{CA'}_{~~~mi}
n_{DB'} \ol\rchi^D e^{BB'}_{~~~k}  { \pt \over \pt \ol \rchi^B}
{\pt \over \pt \psi^C_{~m}}
\cr
&
- {i \over \sqrt{2}} \hbar^2 h^{{3\over 2}}
(e^{CA'i} n_{A}^{~~B'} - e_{A}^{~~B'i} n^{CA'})
n_{DB'} \ol \rchi^D D^P_{~~A'mi}
{ \pt \over \pt \ol \rchi^C} {\pt \over \pt \psi^P_{~~m}},
& (3.11)\cr
}
$$
where the terms containing no matter fields are
consistent with ref. [6,7,14,15,16]. Notice that  a
constant analytical potential is similar to a
cosmological constant term as in  [14,15,16].

The two above expressions will then be used together
with (2.9), (2.12), to obtain the equations for the
bosonic amplitudes coefficients on the wave function
of the universe, $\Psi$. More precisely, we
will employ the integrated form of the
constraints in (2.4), i.e.,
${\cal H} \equiv \int d^3 x H $ obtained from the
action re-written in a canonical  form  as
${\cal S} = \int d^3 x (p\dot{q} - H)$. We stress
 this point in order to deal
correctly with terms as $\pi_\phi \ol \chi_A$ and
hermitian conjugates and to obtain agreement with
known expressions in dimensional-reduced models
[9-12,18,21,25].
There is also another  issue concerning the transformations
(2.12) {\it imposed} by the matter fields
$\chi, \ol \chi$ that should be pointed out. Similarly
to ref.[15,16] we could try to analyse the case
of FRW  with complex scalar fields and fermionic
partners using the Ansatz $\psi^A_{~~i} =
e^{AA'}_{~~~i} \ol \psi_{A'}$ [9-12,18,21,25].
This implies  (see eq. (3.16a)) that
$\beta^A = {3 \over 4} n^{AA'} \ol\psi_{A'} \sim \ol\psi^A$.
However, the analysis in ref. [25]
(see also [9-12,21])
from a one-dimensional reduced action point of view
clearly indicates   that we are required as well
to redefine a new spin-$\half$ component of the
gravitino by $\ol\psi_A \rightarrow h^{{1 \over 4}} \ol\psi_A$.
Only with that we can properly obtain simple
equations for the bosonic amplitudes. Consistently,
a similar procedure has to be employed in this paper  as it
is easy to check. Hence, we have to consider a similar redefinition
as (2.12) to the irreducible spin$ \half, {3 \over 2}$
components of $\psi^A_{~~i}$. Such replacement
is expected to give  the correct supersymmetry constraints in the
$k=1$ FRW model when it is quantized
using the dimensional-reduction alternative approach
via an adequate homogeneous Ans\"atze [9-12,21,25].
Next, we address the construction of a Lorentz
invariant Ansatz for $\Psi$, on which the
supersymmetry constraints derived above will act.

The constraints $ J^{A B} \Psi = 0,~~\bar J^{A' B'} \Psi = 0 $
imply that $
\Psi  $ ought to be a Lorentz-invariant function.
One takes  then expressions in which all spinor indices have been
contracted together. It is reasonable also to consider
only wave functions $ \Psi $ which are spatial scalars, where all spatial
indices $ i, j, \ldots $ have also been contracted together.
 To specify this, note the
decomposition [6,7,14,15,16] of $ \psi^A_{~~B B'} = e_{B B'}^{~~~~i}
\psi^A_{~~i} $:
$$ \psi_{A B B'} = - 2 n^C_{~~B'} \rga_{A B  C} + {2 \over 3} \( \beta_A
n_{B B'} + \beta_B n_{A B'} \) - 2 \eps_{A B} n^C_{~~B'} \beta_C~, \eqno
(3.12) $$
where $ \rga_{A B C} = \rga_{(A B C)} $ is totally symmetric and $
\eps_{A B} $ is the alternating spinor.

As pointed out in the Introduction, we decide to construct our
Lorentz invariant wave function (expressed in several
fermionic sectors and corresponding bosonic amplitudes)
using the framework represented in refs.
[6-12,14-16,18,21,25]. We are aware
of its limitations (they may be particularly severe in
the case of $P(\phi)\neq 0$), as far as the middle sectors are
concerned. In fact, we will be neglecting Lorentz
invariants built with not only
the spin $\half$ and $3 \over 2$
Lorentz irreducible mode components of the fermionic fields
but also with the ones corresponding to the
gravitational degrees of freedom. The new method
proposed in [23,24] do clarify some doubts and paradoxical
situations in supersymmetric quantum cosmology, in particular
constructing the correct middle fermionic sectors.
However, the solutions
in [23,24] for the middle sectors are not entirely new,
as explained in the Introduction. They
were allready present in the ``old''
framework [8]. Thus,
a simpler Lorentz invariant construction
could still be of some utility,
namely in order to
accomodate the basic features of our model and to obtain
new solutions. We hope these will correspond to states
existing in ordinary minisuperspace quantum cosmology.
Nevertheless,
due care must
be taken. In particular,  if no states are
allowed when $P(\phi) \neq 0$. The
proper and complete treatment of our model would then have to
 follow ref.[23,24].  We may expect, however, that our results
could be taken either as a complement or indication
towards its implementation.

Our  general Lorentz-invariant wave
function is then taken to be
a polynomial of eight degree in Grassmann variables
$$\Psi(a_1,a_2,a_3, \phi, \overline{\phi}) = A +
B_1\beta_A\beta^A + B_2\overline{\chi}_A \overline{\chi}^A$$
$$ + C_1 \gamma_{ABC}\gamma^{ABC}$$
$$+ D_1 \beta_A\beta^A \gamma_{EBC}\gamma^{EBC} + D_2 \overline{\chi}_A
\overline{\chi}^A
\gamma_{EBC}\gamma^{EBC}$$
$$+ E_1 (\gamma_{ABC}\gamma^{ABC})^2$$
$$F_1 (\beta_A\beta^A \gamma_{EBC}\gamma^{EBC})^2 + F_2 \overline{\chi}_A
\overline{\chi}^A
(\gamma_{EBC}\gamma^{EBC})^2$$
$$+ G_1 \beta_A\beta^A \overline{\chi}_B \overline{\chi}^B +
H_1 \beta_A\beta^A \overline{\chi}_B \overline{\chi}^B
\gamma_{EDC}\gamma^{EDC}$$
$$+ I_1 \beta_A\beta^A \overline{\chi}_B \overline{\chi}^B
(\gamma_{EDC}\gamma^{EDC})^2$$
$$+ Z_1 \overline{\chi}_A\beta^A + Z_2 \overline{\chi}_A\beta^A
\gamma_{EDC}\gamma^{EDC}$$
$$ + Z_3 \overline{\chi}_A\beta^A (\gamma_{EDC}\gamma^{EDC})^2~.
\eqno(3.13)$$
Note that the term $ \( \beta^A \rga_{A B C} \)^2 = \beta^A
\rga_{A B C} \beta^D \rga_D^{~~B C} $ can be rewritten, using the
anti-commutation of the $ \beta $'s and $ \rga $'s, as
$$ {\rm const.}~\beta^E \beta_E \eps^{A D} \rga_{A B C} \rga_D^{~~B C}
\sim
{}~ \( \beta_E \beta^E \)~\(\rga_{A B C} \rga^{A B C} \).~\eqno(3.14) $$
Similarly, any quartic in $ \rga_{A B C} $ can be rewritten as a multiple of
$ \( \rga_{A B C} \rga^{A B C} \)^2 $. Since there are only four independent
components of $ \rga_{A B C} = \rga_{(A B C)} $, only one independent quartic
can be made from $ \rga_{A B C} $, and it is sufficient to check that $ \(
\rga_{A B C} \rga^{A B C} \)^2 $ is non-zero. Now $ \rga_{A B C} \rga^{A B C}
= 2
\rga_{000} \rga_{111} - 6 \rga_{100} \rga_{011} $. Hence $ \( \rga_{A B C}
\rga^{A B C} \)^2 $ includes a non-zero quartic term $\rga_{000}
\rga_{100} \rga_{110} \rga_{111} $.
The analytic potential imply
that there is coupling between different fermionic levels.

Regarding the ordering chosen for the quantum
mechanical operator form of the supersymmetry
constraints (3.10), (3.11)  let us point  the following.
We decided to adopt the choice made in ref. [17,18], namely
that we order each term cubic in fermions in
$\ol S_{A'}$ (using anti-commutation) such that
one fermionic derivative (momentum) is
on the right and the fermionic variables are on the left.
The ordering of the $S_A$ constraint is
defined
by taking the
hermitian adjoint with respect to the natural inner
product [1,9-12,61];
the terms in $S_A$
cubic in fermions have two derivatives (momenta) on the
right and one fermionic variable (coordinate) on the left.
We stress that the  expressions (3.13), (3.14)
still correspond  to a full theory
formulation (cf. ref. [6,7,14-18]). We could follow
a different path, using  a reduction through homogeneous Ans\"atze
to get an action for a one
one-dimensional time dependent model.
In this context,  one could choose an ordering as in
[9-12], where all
derivatives on the constraint corresponding to our $S_A$
is ordered with all derivatives on the left. In ref.
[11] this ambiguity was taken into account and
an expression for the supersymmetry constraints was
defined,
using again  the
hermitian adjoint with respect to the natural inner
product [1,61]. Moreover, it   related it to the previous one
by putting extra linear tems in the fermionic  coordinate and momenta.
In [25] both the last two approaches were analysed for the
case of a FRW model with supermatter and shown to
be in agreement. We believe that our choice of
ordering
will produce equivalent results to those one could have
obtained had we used a dimensional reduction  from the start
and  possible different  orderings. We will comment on this
 in the following and in section IV.

The action of the constraints operators
 $S_A, \bar S_{A'}$  (3.10), (3.11) on
$\Psi$ (3.13) leads to a system of coupled first order
differential equations which the bosonic amplitude
coefficients of $\Psi$
must satisfy. These coefficients are functions of
$a_1,a_2,a_3,\phi,\bar \phi$. The equations are obtained
after eliminating the $e_i^{AA'}$ and $n^{AA'}$
resulting in the $S_A \Psi = 0, \bar S_{A'} \Psi = 0$
by contracting them with combinations of
$e_j^{BB'}$ and $n^{CC'}$, following by integraton over
$S^3$. These equations correspond essentially to
expressions in front of terms such as
$\ol\chi,\beta,\gamma,\beta^A \ol\chi\ol\chi, \gamma^{ABC} \ol\chi\ol\chi$,
etc,
after the fermionic derivatives in  $S_A, \bar S_{A'}$
have been performed on $S_A \Psi = 0, \bar S_{A'} \Psi = 0$.
As one can easily see, the number of obtained equations will
be very large. Actually, its number will be $44 \times 3$,
taking into account cyclic permutations on $a_1,a_2,a_3$
(see ref. [15,16]). Their full analysis is quite tedious
and to write all the terms just would overburden the reader.
Let us then instead describe  the several steps
involved in the calculations, showing as well some examples
of the calculations involved in solving the
$S_A \Psi = 0, \bar S_{A'} \Psi = 0$ quantum constraints.

The supersymmetry constraints can be described as a
combination of terms constituted by expressions in
$e_i^{AA'}$, $n^{AA'}$ and $\phi, \bar\phi$, with fermionic
variables. More precisely, the $\bar S_{A'}$ constraint
has fermionic terms of the following type:
$\beta^A,\gamma^{ABC}$, $\bar\chi^A,
{\pt \over \pt  \psi^A_i},
{\pt \over \pt \ol \chi^A}$,
$\bar\chi\bar\chi  {\pt \over \pt \ol \chi}$ ,
$\psi\bar\chi
{\pt \over \pt \psi}$ ,
$\psi\bar\chi{\pt \over \pt \ol \chi}$.
The $S_A$ constraint is of second order in fermionic
derivatives and includes  terms as:
${\pt \over \pt \psi^A_i},
{\pt \over \pt \ol \chi^A},
\beta^A, \gamma^{ABC},
\bar\chi^A,
\bar\chi {\pt \over \pt \ol \chi}
{\pt \over \pt \ol \chi},
\psi{\pt \over \pt \ol \chi} {\pt \over \pt \psi},
\bar\chi {\pt \over \pt \ol \chi}
{\pt \over \pt  \psi}$.
In some of the expessions above
we have  deliberately not  written some of the spatial or
spinorial indexes as to allow for their possible combinations
and contractions as one can see from the equations
for $S_A\Psi=0,\bar S_{A'}\Psi=0$. We will use
the gravitino field written in terms of the $\beta$ spin$ \half$ and
$\gamma$  spin $3 \over 2$ modes, respectively.
Moreover, we will use the following expressions:

$$ {\pt (\beta_A\beta^A)\over \pt \psi^B_{~~i}  }=  -
n_A^{~~B'} e_{B B'}^{~~~~i} \beta^A, \eqno (3.15a) $$
$$ {\pt \( \rga_{A D C} \rga^{A D C}
\) \over \pt \psi^B_{~~i}} = - 2 \rga_{B D C}~
n^{C C'} e^{D~~~i}_{~~C'}, \eqno(3.15b) $$
$$ {\pt (\beta^A\beta_A)\over\pt \beta^C} = 2 \beta^C,
\eqno(3.15c) $$
$$ {\pt \beta_A \over \pt \psi^B_{~~j} } =
- \half n_A^{~~B'} e_{BB'}^{~~~j}, \eqno(3.15d)$$
$$ {\pt \gamma^{ADC} \over \pt \psi^B_{~~j}}  =
{1\over 3} n^{CC'} e^{D~~~j}_{~~C'} \epsilon_B^{~~A}
+
{1\over 3} n^{AC'} e^{C~~~j}_{~~C'} \epsilon_B^{~~D}
+
{1\over 3} n^{DC'} e^{A~~~j}_{~~C'} \epsilon_B^{~~C}.
\eqno(3.15e)$$
We also write out
$ \beta^A $ and $ \rga_{B D C} $ in terms of $ e^{E E'}_{~~~~j} $
and $ \psi^E_{~~j} $ as
$$ \beta_A = -\half n_A^{A'i} e_{BA'}^i \psi^B_i, \eqno(3.16a)$$
$$ \gamma_{ABC} =
{1 \over 3} n_C^{~C'} e_{BC'}^{~~~i} \psi_{Ai} +
{1 \over 3} n_A^{~C'} e_{CC'}^{~~~i} \psi_{Bi}
+ {1 \over 3} n_B^{~C'} e_{AC'}^{~~~i} \psi_{Ci}, \eqno(3.16b) $$
from which (3.15a), (3.15b) become
$$
{\pt \(\beta_A\beta^A\) \over \pt \psi^B_i} =
{1 \over 4} e_{BB'}^i e_C^{B'j} \psi^C_j,
\eqno(3.17a)
$$
$$
\eqalignno{  {\pt \(\gamma_{ADC} \gamma^{ADC} \) \over
\pt \psi^B_i} & = - {1 \over 3} e_{DA'}^j e^{DA'i}
\psi_{Bj} \cr
& - {2 \over 3} n_B^{~A'} n^{CC'} e_{CA'}^{~~~j} e^{D~~i}_{~~C'}
\psi_{Dj}
+ {1 \over 3} e_{BA'}^{~~~j} e^{CA'i} \psi_{Cj}. &(3.17b) \cr}$$

Since the wave function is of even order in fermionic variables,
the equations  $S_A\Psi=0, \bar S_{A'}\Psi=0$ will be of odd order
in fermionic variables, up to seventh order. Notice that
some of the fermionic terms in (3.10), (3.11)  applied to $\Psi$
increase the fermionic order by a factor of one (e.g, $\bar\chi$)
while others as $\bar\chi {\pt \over \pt \ol \chi}
{\pt \over \pt \psi}$ decrease
it by the same amount.

The two following tables illustrate in a simple
way how the quantum
supersymmetry constraints (3.10), (3.11) operate on $\Psi $(3.13)
and which types
of fermionic terms can be obtained.
The first table correspond to $S_A \Psi$ and the second
to $\ol S_{A'} \Psi$.
The first row is constituted by the fermionic
operators present in the supersymmetry constraints
while in the first column we have the several terms which are
consequently obtained from $S_A\Psi=0,\bar S_{A'}\Psi=0$.
In the intersection slots we have the possible different bosonic
coefficients of the wave function $\Psi$, representing
the specific way  a fermionic operator acts on $\Psi$.
For example, we see that the operator
${\ol \chi}$ in $\bar S_{A'}$  produces fermionic
terms such as
$\beta\ol\chi\ol\chi$, $\beta\ol\chi\ol\chi\gamma\gamma$,
$\beta\ol\chi\ol\chi(\gamma\gamma)^2$
to which correspond
bosonic functions $Z_1, Z_2, Z_3$, respectively. In such a way
we can infer which type of wave function coefficients
are present in each equation (say, the one corresponding
to $\ol\chi\beta\beta(\gamma\gamma)^2$)
and which
fermionic operators have produced those terms.
The brackets
in $(\gamma\gamma)$ stand for $\gamma^{ABC}\gamma_{ABC}$
and the same notation will be used for the other
fermionic variables throughout the paper.
A slash over a particular bosonic amplitude
means that after all the calculations have been
made to simplify the corresponding equations, the
expression associated with that particular coefficient
is zero. An example is when those expressions
contain something like $\gamma^{ABC}\epsilon_{BC}$ which
is zero due to the symmetry of $\gamma^{ABC} = \gamma^{(ABC)}$ and
the antisymmetry of $\epsilon_{AB}=\epsilon_{[AB]}$.
Hence, they do not contribute to the set
of equations we will discuss in the following two subsections.
A ``$\bullet$'' means that no bosonic
amplitudes in $\Psi$ can match the particular
fermionic
operator and the term in $S_A \Psi = 0, \ol S_{A'} \Psi = 0$,
respectively.
As a last comment, the third and fourth columns in both
tables corespond to the
action of fermionic operators which appear in terms
involving $P(\phi), D_\phi P(\phi)$ respectively, or their
hermitian conjugates.

\bigskip

\bigskip

\input tables


\begintable
$\searrow$ \| ${\pt \over \pt \psi}$ | ${\pt \over \pt \ol \chi}$ \|
$\psi$ | $\ol\chi$ \| $\ol\chi{\pt \over \pt \ol\chi}
{\pt \over \pt \ol\chi}$ | $\psi
{\pt \over \pt \ol \chi}
{\pt \over \pt \psi}$
| $\ol\chi
{\pt \over \pt \ol \chi}
{\pt \over \pt \psi}$
\crthick
$\beta$            \| $B_1$  | $Z_1$         \| $A$  | $\bullet$   \| $\bullet$
   | $Z_1$  | $\bullet$
\cr
$\gamma$           \| $C_1$  | $\bullet$     \| $\A$ | $\bullet$   \| $\bullet$
   | $\Z_1$ | $\bullet$
\cr
$\ol\chi$          \| $Z_1$  | $B_2$         \| $\bullet$ |     $A$ \| $B_2$
 |  $\bullet$  | $Z_1$
\cr
$\beta\beta\ol\chi$\| $\bullet$ | $G_1$      \| $Z_1$    |$B_1$   \| $\G_1$
 | $\G_1$    | $\bullet$
\cr
$\beta\beta\gamma$ \| $D_1$  | $\bullet$     \| $\B_1$  | $\bullet$ \|
$\bullet$  | $\Z_2$   | $\bullet$
\cr
$\beta\ol\chi
\ol\chi$           \| $G_1$  | $\bullet$     \| $B_2$| $Z_1$       \| $\bullet$
   | $\bullet$ | $\G_1$
\cr
$\beta\ol\chi
\gamma$            \| $Z_2$  | $\bullet$     \| $\bullet$ | $\bullet$ \|
$\bullet$ | $D_2,G_1$  | $Z_2$
\cr
$\beta\beta\gamma$ \| $D_1$  | $Z_2$        \| $C_1$  | $\bullet$   \|
$\bullet$    | $Z_2$  | $\bullet$
\cr
$\ol\chi\ol\chi
\gamma$            \| $D_2$  | $\bullet$    \| $\B_2$  | $\bullet$   \|
$\bullet$    | $\bullet$ | $\D_2$
\cr
$\ol\chi\gamma
\gamma$             \| $Z_2$  | $D_2$       \| $\bullet$ | $C_1$    \| $D_2$
     | $D_2$     | $\bullet$
\cr
$\gamma\gamma
\gamma$            \| $E_1$   | $\bullet$   \| $\C_1$    | $\bullet$ \|
$\bullet$    | $\Z_2$ |   $\bullet$
\cr
$\beta(\gamma
\gamma)^2$          \|$F_1$   | $Z_3$ \| $E_1$| $\bullet$   \| $\bullet$    |
$Z_3$ | $\bullet$
\cr
$\gamma\beta\beta
\gamma\gamma$      \|$F_1$   |$\bullet$      \|$\D_1$|  $\bullet$  \| $\bullet$
   |$\Z_3$| $\bullet$
\cr
$\gamma\ol\chi\ol
\chi\gamma
\gamma $           \|$F_2$   | $\bullet$     \|$\D_2$| $\bullet$   \| $\bullet$
   | $\bullet$    | $\F_2$
\cr
$\beta\ol\chi
\ol\chi
\gamma\gamma$     \|$H_1$    |$\bullet$     \| $D_2$ |$Z_2$\|  $\bullet$   |
$\bullet$  |$\H_1$
\cr
$\gamma\beta
\beta\ol\chi
\ol\chi$          \|$H_1$    |  $\bullet$           \|$\G_1$  |$\bullet$   \|
$\bullet$  | $\bullet$   |$\H_1$
\cr
$\ol\chi
(\gamma\gamma)^2$ \|$Z_3$    | $F_2$\| $\bullet$    |$E_1$\|$F_2$  |$F_2$|$Z_3$
\cr
$\ol\chi
\beta\gamma
\gamma\gamma$     \|$Z_3$    |  $\bullet$   \| $\bullet$    |  $\bullet$  \|
$\bullet$    |$F_2, H_1$|$Z_3$
\cr
$\ol\chi\beta
\beta\gamma
\gamma$           \|  $\bullet$      | $H_1$\|$Z_2$ | $\bullet$
\|$\H_1$|$\H_1$| $\bullet$
\cr
$\beta\ol\chi\ol
\chi\(\gamma
\gamma\)^2$        \| $I_1$   | $\bullet$    \|$F_2$ |$Z_3$\|  $\bullet$  |
$\bullet$     |$\I_1$
\cr
$\gamma
\beta\beta
\ol\chi
\ol\chi
\gamma\gamma$     \| $I_1$   |  $\bullet$   \|  $\H_1$ | $\bullet$  \|
$\bullet$   |  $\bullet$   |$\I_1$
\cr
$\ol\chi\beta
\beta(\gamma
\gamma)^2$        \| $\bullet$       | $I_1$\|  $Z_3$ |   $F_1$  \|  $\I_1$|
$\I_1$   |    $\bullet$
\endtable

\centerline{{\bf Table 1}: Action of $S_A$ (3.11) on $\Psi$ (3.13)}

\begintable
$\searrow$ \| $ \psi$ | $ \ol \chi$ \|
${\pt \over \pt \psi}$ | ${\pt \over \pt \ol\chi}$ \|
$\ol\chi \ol\chi
{\pt \over \pt \ol\chi}$ | $\psi
 \ol \chi
{\pt \over \pt \psi}$
| $\psi \ol\chi
{\pt \over \pt \ol \chi}$
\crthick
$\beta$            \| $A$  | $\bullet$ \| $B_1$  | $Z_1$   \| $\bullet$   |
$\bullet$  | $\bullet$
\cr
$\gamma$           \| $A$  | $\bullet$     \| $\C_1$ | $\bullet$   \| $\bullet$
   | $\bullet$ | $\bullet$
\cr
$\ol\chi$          \| $\bullet$  | $A$ \| $Z_1$    |  $B_2$\| $\bullet$ |
$\bullet$     | $\bullet$
\cr
$\beta\beta\ol\chi$\| $Z_1$   | $B_1$ \| $\bullet$ |$G_1$ \| $\bullet$    |
$B_1$     | $Z_1$
\cr
$\beta\beta\gamma$ \| $B_1$  | $\bullet$     \| $\D_1$| $\bullet$   \|
$\bullet$    | $\bullet$ | $\bullet$
\cr
$\beta\ol\chi
\ol\chi$           \| $B_2$  | $Z_1$ \| $G_1$| $\bullet$\| $Z_1$   | $Z_1$ |
$B_2$
\cr
$\beta\ol\chi
\gamma$            \| $Z_1$  | $\bullet$     \| $\bullet$    | $\bullet$   \|
$\bullet$    |$C_1,B_1$| $Z_1$
\cr
 $\beta\gamma\gamma$\| $C_1$  | $\bullet$ \| $D_1$| $Z_2$   \| $\bullet$    |
$\bullet$  | $\bullet$
\cr
$\ol\chi\ol\chi
\gamma$            \| $B_2$  | $\bullet$     \|$\D_2$| $\bullet$   \| $\bullet$
   | $\Z_1$     |$\B_2$
\cr
$\ol\chi\gamma
\gamma$             \| $\bullet$  | $C_1$  \| $Z_2$    |  $D_2$ \| $\bullet$ |
$C_1$  | $\bullet$
\cr
$\gamma\gamma
\gamma$            \|$C_1$   |  $\bullet$    \|$\E_1$| $\bullet$   \| $\bullet$
   |$\bullet$ |   $\bullet$
\cr
$\beta(\gamma
\gamma)^2$          \|$E_1$   | $\bullet$  \| $F_1$| $Z_3$   \| $\bullet$    |
$\bullet$ | $\bullet$
\cr
$\gamma\beta\beta
\gamma\gamma$      \|$D_1$   |$\bullet$      \|$\F_1$|  $\bullet$  \| $\bullet$
   | $\bullet$ |$\bullet$
\cr
$\gamma\ol\chi\ol
\chi\gamma
\gamma $           \|$D_2$   | $\bullet$     \|$\F_2$| $\bullet$   \| $\bullet$
   | $Z_2$   | $\D_2$
\cr
$\beta\ol\chi
\ol\chi
\gamma\gamma$     \|$D_2$    |$Z_2$ \| $H_1$ |$\bullet$ \|  $Z_2$   | $Z_2$
|$D_2$
\cr
$\gamma\beta
\beta\ol\chi
\ol\chi$          \|$G_1$    |  $\bullet$   \|$\H_1$ |$\bullet$   \|  $\bullet$
   | $Z_2$  |$\H_1$
\cr
$\ol\chi
(\gamma\gamma)^2$ \| $\bullet$    | $E_1$ \| $Z_3$ | $F_2$ \|  $\bullet$
|$E_1$|  $\bullet$
\cr
$\ol\chi
\beta\gamma
\gamma\gamma$     \|$Z_2$    |  $\bullet$   \| $\bullet$    |  $\bullet$  \|
$\bullet$    |$D_1,E_1$|$Z_2$
\cr
$\ol\chi\beta
\beta\gamma
\gamma$           \|  $Z_2$      | $D_1$\| $\bullet$    | $H_1$ \| $\bullet$ |
$D_1$| $Z_2$
\cr
$\beta\ol\chi\ol
\chi\(\gamma
\gamma\)^2$        \| $F_2$   | $Z_3$\|$I_1$ | $\bullet$ \|  $Z_3$  | $Z_3$
|$F_2$
\cr
$\gamma
\beta\beta
\ol\chi
\ol\chi
\gamma\gamma$     \| $H_1$   |  $\bullet$   \|$\I_1$ | $\bullet$  \|
$\bullet$   |  $\bullet$   |$\H_1$
\cr
$\ol\chi\beta
\beta(\gamma
\gamma)^2$        \| $Z_3$       | $F_1$\| $\bullet$    |$I_1$\| $\bullet$
| $F_1$|$Z_3$
\endtable

\centerline{{\bf Table 2}: Action of $\ol S_{A'}$ (3.10) on $\Psi$ (3.13)}

It is worthwhile to stress the following important
property, which holds regardless we put $P(\phi)=0$ or not.
Using the symmetry properties of $e^{AA'}_i, n_{AA'},
\gamma^{ABC}, \epsilon_{AB}$ we can check that all
equations which correspond to the terms
$\gamma, \gamma\beta\beta, \ol\chi\ol\chi\gamma,
\gamma\gamma\gamma$ , $\gamma\beta\beta\gamma\gamma$,
$\gamma\ol\chi\ol\chi\gamma\gamma$,
$\gamma\beta\beta\ol\chi\ol\chi$,
$\gamma\beta\beta\ol\chi\ol\chi\gamma\gamma$
in $S_A\Psi = 0, \bar S_{A'}\Psi=0$
will give a similar expression for
the coefficients $A,B_1, B_2, C_1,$ $ D_1,$ $D_2$, $E_1$,
$F_1$, $ F_2$, $G_1$, $H_1$, $I_1$.
Namely,
$$ P(a_1,a_2,a_3;\phi,\ol\phi) e^{\pm\(a_1^2 + a_2^2 + a_3^2\)}$$
(cf.ref. [7,8,23,24] and see below). The same does not apply
to the $Z_1, Z_2, Z_3$ coefficients as the $\beta\ol\chi\gamma$ and
$\beta\ol\chi\gamma(\gamma\gamma)$ terms
from both  the supersymmetry constraints just mix them  with
other coefficients in $\Psi$.

In the following we will analyse two cases separately: when
the analytic potential $P(\Phi)$ is arbitrarly and
when is identically set to zero. We
will begin by the former.

\vskip 1cm

\centerline{\bf Case $\cal A$. $P(\phi)\neq 0$}

\vskip 1cm

 From the equations corresponding to
$\gamma_{DEF}(\beta \beta)$ and $\gamma_{FGH}(\gamma\gamma)$
in $\ol S_{A'} \Psi = 0$ we get
$$ 2 \eps^{i j k} e_{A A' i} \om^A_{~~B j} n^D_{~~B'} e^{C B'}_{~~~~k}
B_{1}
- \hbar  n^D_{~~B'} e^{C B'}_{~~~~i} {\de B_{1} \over \de e^{B
A'}_{~~~~i}} $$
$$ + \( B C D \to C D B \) + \( B C D \to D B C \) = 0~, \eqno (3.18) $$
$$ 2 \eps^{i j k} e_{A A' i} \om^A_{~~B j} n^D_{~~B'} e^{C B'}_{~~~~k}
C_{1}
- \hbar  n^D_{~~B'} e^{C B'}_{~~~~i} {\de C_{1} \over \de e^{B
A'}_{~~~~i}} $$
$$ + \( B C D \to C D B \) + \( B C D \to D B C \) = 0~. \eqno (3.19) $$
Contracting Eq.~(3.18), (3.19)  with $ e^{B A' \ell} n_{C C'} e_D^{~~C' N} $,
multiplying by, say,
$$ \de h_{\ell n} = {\pt h_{\ell n} \over \pt a_1} = 2 a_1 E^1_iE^1_j, $$
and integrating over $S^3$
gives
$$ 3 \hbar a_1 {\pt B_1 \over \pt a_1} - \hbar  \(
a_1
{ \pt B_1 \over \pt a_1} + a_2 {\pt B_1
\over \pt a_2} + a_3 {\pt B_1 \over \pt a_3} \) $$
$$ - 16 \pi^2 a_1 a_2 a_3 \( {a_3 \over a_1  a_2} +
{a_2 \over a_3 a_1} - 2 {a_1 \over a_2 a_3} \)
B_1
= 0~,
\eqno (3.20) $$
$$ 3 \hbar a_1 {\pt C_1 \over \pt a_1} - \hbar \kap^2 \(
a_1
{ \pt C_1 \over \pt a_1} + a_2 {\pt C_1
\over \pt a_2} + a_3 {\pt C_1 \over \pt a_3} \) $$
$$ - 16 \pi^2 a_1 a_2 a_3 \( {a_3 \over a_1  a_2} +
{a_2 \over a_3 a_1} - 2 {a_1 \over a_2 a_3} \)
C_1
= 0~,
\eqno (3.21) $$
and the corresponding  equations given by permuting $
a_1 a_2 a_3 $ cyclically.

Following the steps in ref. [14,15,16], we consider
now eq. (3.20), (3.21) and cyclic permutations.
These leads to
$$ \hbar  \( a_1 {\pt B_1 \over \pt a_1} -
a_2  {\pt B_1 \over \pt
a_2} \) = 16 \pi^2 \( a_2^2 - a_1^2 \) B_1 \eqno (3.22) $$
$$ \hbar  \( a_1 {\pt C_1 \over \pt a_1} -
a_2  {\pt C_1 \over \pt
a_2} \) = 16 \pi^2 \( a_2^2 - a_1^2 \) C_1 \eqno (3.23) $$
and cyclic permutations.
Integrating them   along a
characteristic $ a_1 a_2 = $ const., $ a_3 = $ const., using the parametric
description $ a_1 = u_1 e^\tau $, $ a_2 = u_2 e^{- \tau} $,
we get
$$ B_1 = f (a_1 a_2, a_3; \phi, \ol\phi) e^{ - {8 \pi^2 \over \hbar}
{}~ \( a_1^2 + a_2^2 \) }~, \eqno (3.24) $$
and similarly to $C_1$.
 From their cyclic permutations and from the
invariance under this cyclic permutation
over $a_1, a_2, a_3$ we subsquently obtain
$$ B_1 = f (a_1 a_2 a_3;\phi,\ol\phi) e^{ - {8 \pi^2 \over \hbar} ~
\( a_1^2 + a_2^2 + a_3^2 \) }~, \eqno (3.25) $$
$$ C_1 = g (a_1 a_2 a_3; \phi, \ol\phi) e^{ - {8 \pi^2 \over \hbar}
{}~\( a_1^2 + a_2^2 + a_3^2 \) }~, \eqno (3.26) $$
The same type of expressions follow for the remaining
bosonic coefficients with the
exception of $Z_1, Z_2, Z_3$.

Let us then proceed considering the equations obtained from
$S_A \Psi = 0$ and first order in fermions,
with terms linear in $\beta$ and $\gamma$.
These equations (see also next subsection),
after contraction with expressions in
$e^{AA'}_i, n_{AA'}$ and integrating over  $S^3$ and
 using
 (3.15)-(3.17), can be combined into
$$ \eqalignno {
&{1 \over 16} \hbar^2  \( a_1 {\pt B_1 \over \pt a_1} + a_2
{\pt B_1 \over \pt a_2} + a_3 {\pt B_1 \over \pt a_3} \) \cr
- &{1 \over 3} \hbar  a_1 \[ 3 {\pt C_1 \over \pt a_1}
- a_1^{- 1} \( a_1
{\pt C_1 \over \pt a_1} + a_2 {\pt C_1 \over \pt a_2} + a_3 {\pt
C_1 \over \pt a_3} \) \] \cr
&- 16 \pi^2 e^{K/2} P(\phi) a_1 a_2 a_3 A
- \pi^2 \hbar a_1 a_2 a_3  \( {a_1 \over a_2 a_3} +
{a_2 \over a_3 a_1} + {a_3  \over a_1 a_2} \)
B_1 \cr
&+ {1 \over 3} \( 16 \pi^2 \) \hbar
a_1 a_2 a_3  \( {2 a_1 \over a_2 a_3} - {a_2 \over a_3 a_1} -
{a_3  \over a_1 a_2} \) C_1  \cr
&
+ 8\pi^2 \ol\phi \hbar^2 Z_1 - 2i \hbar^2 {\pt Z_1 \over \pt \phi}
=0
{}~, &(3.27) \cr
} $$
and two more equations given by cyclic permutation of $ a_1 a_2 a_3 $.
Let us point out that the $\beta$ equation  correspond to
the trace of (3.27)  while the $\gamma$ one
represents its trace free part.
Eq.  (3.27) differs from the one in ref. [15,16]
in its last two terms.
 From eq. (3.25), (3.26) and (3.27) with cyclic
permutations we get
$$ \eqalignno {
&16 \pi^2 e^{K/2} P(\phi) A
= - 2 \pi^2 \hbar (a_1 a_2 a_3)^{- 1} \( a_1^2 + a_2^2 + a_3^2 \)
e^{ \[ - {8 \pi^2 \over \hbar}~\( a_1^2
+ a_2^2 + a_3^2 \) \]}
 f \cr
&+ {3 \over 16} \hbar^2 \kap^2
e^{ \[ - {8 \pi^2 \over \hbar}~\( a_1^2
+ a_2^2 + a_3^2 \) \]}
f^{\prime} \cr &
+ {2 \over 3} \( 16 \pi^2 \) \hbar
(a_1 a_2 a_3)^{- 1} \( 2 a_1^2 - a_2^2 - a_3^2
\)~
e^{ \[ - {8 \pi^2 \over \hbar}~\( a_1^2
+ a_2^2 + a_3^2 \) \]}
 g  \cr
 &
 + 8 \pi^2 \hbar^2 (a_1 a_2 a_3)^{-1} \phi Z_1
 - 2i \hbar^2 (a_1 a_2 a_3)^{-1} {\pt Z_1 \over
 \pt  \phi}
 &(3.28) \cr } $$
and cyclically.
We assume as well as in [14,15,16]
that
the coefficients in the the wave function are
invariant under permutations of $a_1,a_2,a_3$.
 We then get  $g=0 \Rightarrow C_1 = 0$ as the only
 possible solution.

Now, eq. (3.28) and its cyclic
permutations with $C_1=0$ must be solved consistently
with the equation obtained from the linear terms in
$\beta$ and $\gamma$ from $\bar S_{A'} \Psi = 0$. We follow
the same procedure as above, writting $\beta$ and $\gamma$
in terms of $\psi_A^{~~i}$, adding them and geting
$$ \hbar  a_1 {\pt A \over \pt a_1} + 1 6 \pi^2 a_1^2
A + 6 \pi^2 \hbar
e^{K/2} P(\phi) a_1 a_2 a_3 B_1
- \sqrt{2} 16 \pi^2 \hbar a_1 a_2 a_3 e^{K/2} (D_\phi P(\phi)) Z_1
= 0~, \eqno (3.29) $$
and two others given by cyclic permutation of $ a_1 a_2 a_3$.
Relatively to ref. [14,15,16], the main difference is in the
term in $Z_1$. Eliminating $A$, we get
$$ \eqalignno {
&{3 \hbar^3  \over 16
\( 16 \pi^2 e^{K/2} P(\phi) \)} f'' - {\hbar^2  \over 8
e^{K/2} P(\phi)}~{\( a_1^2 + a_2^2 + a_3^2 \) \over a_1 a_2 a_3} f' \cr
&+ 6 \pi^2 \hbar e^{K/2} P(\phi) f -
{\hbar^2  \over 4 e^{K/2} P(\phi)}~{1 \over a_2^2 a_3^2} f +
{\hbar^2  \over 8 e^{K/2} P(\phi)}
{}~{\( a_1^2 + a_2^2 + a_3^2 \)
\over (a_1 a_2 a_3)^2} f \cr
& -
{\hbar^3 \over 16 \pi^2}  {1 \over e^{K/2} P(\phi)}
{1 \over \(a_1 a_2 a_3\)^2} \ol\phi
e^{ \[  {8 \pi^2 \over \hbar}~\( a_1^2
 a_2^2 + a_3^2 \) \]}
 Z_1 \cr
 &
 {\hbar^2 \over 2 e^{K/2} P(\phi)} \ol\phi
e^{ \[  {8 \pi^2 \over \hbar}~\( a_1^2
 a_2^2 + a_3^2 \) \]}
 {Z_1 \over a_2^2 a_3^2}
 \cr
& + {2i \hbar^3 \over 16\pi^2} {1 \over e^{K/2} P(\phi)}
{1 \over \(a_1 a_2 a_3\)^2}
e^{ \[  {8 \pi^2 \over \hbar}~\( a_1^2
+ a_2^2 + a_3^2 \) \]} {\pt Z_1  \over \pt \phi} \cr
&
- {2i\hbar^2 \over e^{K/2} P(\phi)}
e^{ \[  {8 \pi^2 \over \hbar}~\( a_1^2
 a_2^2 + a_3^2 \) \]}
  {1 \over a_2^2 a_3^2} {\pt Z_1 \over\pt \phi}
\cr
& -
\sqrt{2} 16 \pi^2 i \hbar e^{K/2} D_\phi P(\phi)
e^{ \[  {8 \pi^2 \over \hbar}~\( a_1^2
 a_2^2 + a_3^2 \) \]}
Z_1
= 0~, &(3.30) \cr } $$
and cyclic permutations.
Since $ f = f (a_1 a_2 a_3; \phi,
\ol \phi) $ is invariant under
permutations of $a_1,a_2,a_3$, the
terms in $ (a_2 a_3)^{- 2} f $ term and its permutations imply now,
differently from [14,15,16], a relation between $Z_1$ and $B_1$:
$$
f(a_1 a_2 a_3; \phi, \ol\phi)
e^{ \[  {8 \pi^2 \over \hbar}~\( a_1^2
 a_2^2 + a_3^2 \) \]}  + 2\ol\phi Z_1
 - 8i {\pt Z_1 \over \pt \phi} = 0.\eqno(3.31) $$

For the particular case of $B_1 = 0$,
i.e., ($f = 0$), it follows in the end of the day that from the remaining
equations
 the only possible solution is $\Psi = 0$.
Considering the equations
 from the linear term
in $\beta$ and $\gamma$ from $\bar S_{A'} \Psi = 0$,
the only term different from ref. [14,15,16] is the
one in $Z_1$. But it happens that the solutions of the
equation in $\gamma$ are the same as in ref. [7]
(cf. eq. (3.18)-(3.26)).
So, substituting these solution back in the
equation (3.29) -- see also table 2 for the $\beta$ equation --
 all terms safe the last will contribute to give zero as they
are allready present in the corresponding
equation for the Bianchi-IX model with $\Lambda = 0$ [7].
Hence, we are left out with  $Z_1 = 0$. As a consequence,
the equation corresponding to the term in $S_A \Psi = 0$
linear in $\beta$ gives then $A=0$. Using all these
results in the equation corresponding to the tems linear
in $\ol\chi$ from $\bar S_{A'}\Psi=0$ gives $B_2=0$. Then
the $\beta\beta\ol\chi$ term from $\ol S_{A'}\Psi=0$ implies $G_1=0$.
Consequently the $\gamma\beta\beta\ol\chi\ol\chi$ term
from the $\ol S_{A'}$ constraint
leads to $Z_2=0$. The $\beta\ol\chi\gamma$ term from $S_A$
gives that
$D_2=0$. Again, collecting these results in the $\beta\gamma\gamma$
term from $\bar S_{A'} \Psi = 0$ we obtain $D_1 = 0$.
The $\beta\ol\chi\ol\chi\gamma\gamma$ term from
$\bar S_{A'} \Psi = 0$  leads to $H_1 = 0$ and the
$\beta\ol\chi\gamma \gamma \gamma$ term to $E_1=0$.
Finally, we have to address the coefficients $F_1,F_2,I_1,Z_3$
of $\Psi$. Their analysis turn out to be similar
to the one of $A,B_1,B_2,Z_1$ but without the
$C_1$ coefficient. Using the equations
corresponding to
$\beta\ol\chi\ol\chi(\gamma\gamma)^2$ from $\bar S_{A'} \Psi = 0$,
$\gamma\ol\chi\ol\chi\gamma\gamma$, $\gamma\beta\beta\gamma\gamma$
and $\beta\ol\chi\ol\chi(\gamma\gamma)^2$ terms in $S_A \Psi = 0$
we get that $F_2 = 0$.
Then $\ol\chi(\gamma\gamma)^2$ in $\bar S_{A'}\Psi=0$,
$\beta(\gamma\gamma)^2$ in  $\bar S_{A'}\Psi=0$ and
$\beta\ol\chi\beta(\gamma\gamma)^2$ in $\bar S_{A'}\Psi=0$
induce  that $Z_3=0$, $F_2=0$ and $I_1=0$, respectively.
However, these results are just a consequence of a rather
particular case. We should address then a more general situation.

For an arbitrarly $f$, eq. (3.31) allows to write an
expression for $Z_1$ in terms of functions of
$\phi, \ol\phi$ and $a_1, a_2, a_3$.
If we use that expression in other equations, we
get other formulas for other bosonic coefficients.
For example, from the equation in $\beta$ from $\ol S_{A'}
\Psi = 0$ we may use
$B_1, Z_1$ as above to get an expression for $A$. One
complements these steps with the equation linear in
$\ol \chi$ from $\ol S_{A'} \Psi = 0$ and study
$B_2$. Following this procedure, we would get
the general solution of this extremely complicated
set of differential equations.
Although apparently possible, we could not
establish a definite result in the end
due to the complexity of the equations involved.
As in [10,12], no easy way is apparent of
obtaining an analitycal  solution to this set of
equations. Moreover, the exponential terms
$e^{K/2}$ leads to some difficulties.

We could also speculate by saying that non trivial
physical states could be found for a generic $P(\phi)$.
In fact, the presence of the anisotropic gravitational
degrees of freedom and the spin$ {3 \over 2}$ modes of
the gravitino together with the matter fields produce
some important changes relatively to ref. [14,15,16].
In particular, the structure of the equations is
different as expressed in (3.27)-(3.28) and namely in
(3.31).

Let us also point that for {\it frozen}
scalar fields, i.e., such that
$e^{K/2} P(\phi) \equiv const$ we would obtain a scenario
similar
to an effective cosmological constant term.
In fact, the equations will turn to be  familiar
to the ones in [14,15,16]. Hence, from the
asumption $\phi, \ol\phi = const$ we would be led again to
$\Psi = 0$.

Nevertheless,
 we should
stress that the main (and perhaps, solely) conclusions
from the case $\cal A$ is that the adequacy of Ansatz (3.11) has
severe limitations. Again, it seems paradoxical
that for all degrees of freedom of the Bianchi-IX
(diagonal) model  with supermatter and analytical potential,
the constraints imply that $\Psi = 0$
for some simplified assumptions. The proper answer
to this situation would have to be addressed
from the point of view ref. [23,24]. The Ansatz for $\Psi$
is definitevely incomplete as far as chirality breaking/mode
mixing (non-conservation of fermion number in the sense of
[23,24]) is concerned. The  results in this
subsection can be  be taken as another example
strengthening the arguments of R. Graham and A. Csord\'as [23,24], although
 expression (3.11) may still be useful
to indicate new possible  solutions.

\vskip 1cm

\centerline{\bf Case $\cal B$. $P(\phi) =  0$}

\vskip 1cm

Let us now consider the case when we choose the analytical potential
to be identically zero. The analysis of our Bianchi-IX model
with supermatter will turn out to be simpler then in the previous
subsection. Moreover, we will follow closely some arguments and framework
presented in [9-12,18,21,25].

As we can see either from the equations directly obtained from
$S_A \Psi = 0, \bar S_{A'} \Psi = 0$ or from their
representation in the tables 1 and  2, we have self-contained
groups of equations relating the 15  wave function coefficients
(in the case of $P(\Phi) \neq 0$ that is not true).
This applies to 3 groups involving $(A, B_1, B_2, C_1, Z_1)$,
$(G_1, D_1, D_2, E_1, Z_2)$ and $(H_1, F_1, F_2, I_1, Z_3$).
Moreover,
notice in particular
that the equation corresponding to the terms linear in
$\beta,\gamma,\ol\chi$ in $\bar S_{A'}\psi = 0$ and
$\beta\ol\chi\ol\chi(\gamma\gamma)^2, \beta\beta\ol\chi(\gamma\gamma)^2$
in $S_A \Psi = 0$ completly determine the coefficients
$A$ and $I_1$. Moreover, $A$ and $I_1$  do not appear in any other
equation. We have then
$$A = f(\phi) e^{-{{8\pi^2}\over \hbar} [a_1^2 + a_2^2 + a_3^2]}, \eqno(3.32)$$
$$I_1 = k({\ol\phi}) e^{{{8\pi^2}\over \hbar}[a_1^2 + a_2^2 + a_3^2]}
e^{-2\pi^2\phi
\overline{\phi}}. \eqno(3.33)$$

The coefficients $G_1, H_1$ share a similar property
with respect to the $\beta\beta\ol\chi, \beta\ol\chi\ol\chi$ equations from
$S_A \Psi = 0$ and $\beta\beta\ol\chi\gamma\gamma, \beta\ol\chi\ol\chi
\gamma\gamma$ from $S_A \Psi = 0$, respectively.
However, the $\beta\ol\chi\gamma$ and $\beta\ol\chi\gamma(\gamma\gamma)$
terms in $S_A \Psi = 0$ impose that
$G_1$ also appear together with $Z_2,D_2$ and that $H_1$ is together
with $Z_3,F_2$ in independent expressions.
This particular property will allow to determine
that $G_1$ and $ I_1 $ from $Z_2, D_1$ and $Z_3, F_2$,
respectively.

The equations involving
$B_1,B_2,C_1,Z_1$ can also be said to be self-contained in the
same sense that they involve only these coeficients and
no other. Moreover, these coefficients do not occur in any
other equations. This can be easily checked, namely from the
equations for the terms linear in $\beta,\ol\chi$ in $S_A \Psi = 0$,
$\beta\beta\ol\chi$ and $\beta\ol\chi\ol\chi$ in $\bar S _{A'} \Psi = 0$,
$\beta\beta\gamma$ in $\bar S _{A'} \Psi = 0$,
$\ol\chi\gamma\gamma, \beta\gamma\gamma, \gamma\gamma\gamma$ in
$\bar S _{A'} \Psi = 0$, $\gamma$ in $S_A \Psi = 0$.
The previous ones  in $\ol \chi\gamma\gamma, \beta\gamma\gamma,
\gamma\gamma\gamma, \gamma$ just involve $C_1$.
All the others just have $B_1, B_2, Z_1$.
However, the $\beta\ol\chi\gamma$ equation in $\bar S_{A'} \Psi = 0$
mixes $B_1,C_1,Z_1$. Actually, is the only equation which
mixes $C_1$ with the remaining  bosonic coefficients in the
corresponding group.

The same structure of equations and relations between coefficients
can be easily checked to occur as well to the subsets involving
$D_1,D_2, E_1,Z_2$ and  $F_1,F_2,I_1,Z_3$. Notice again that
for the first group, the equations from $\beta(\gamma\gamma)^2,
\ol\chi(\gamma\gamma)^2$ in $S_A \Psi = 0$ and
$\gamma\gamma\gamma$ in $S_A \Psi = 0$
 involve only $E_1$ and are
enough to determine it. Moreover, the
$\beta\ol\chi\gamma(\gamma\gamma)$ term in
$\bar S_{A'} \Psi = 0$ relates $Z_2,D_1,E_1$.

We now  proceed to analyse the groups of equations
which includes $B_1,B_2,C_1,Z_1$ .
We also include the redefinitions $ B_1 \rightarrow iB,
B_2 \rightarrow B_2$ to simplify the results
(cf. ref.  [10-12,18,25]).
The analysis of  the remaining
groups is similar, as stated above.
The $\beta\beta\ol\chi$, $\beta\ol\chi\ol\chi$    equations
from $\ol S_{A'} \Psi = 0$ and $\beta$, $\ol \chi$
equations from $S_A \Psi = 0$
give, respectively,
$$
{4\hbar^2 }
{ \pt B_1 \over \pt \ol\phi} +
8\pi^2  \phi B_1
+ {3 \hbar^2 \over 4} a_i  {\pt Z_1 \over \pt a_i}
 - {3\hbar^2  \over 4\sqrt{2}} Z_1 +
 4 \pi^2 \hbar
  a_1 a_2 a_3 \( {a_1 \over a_2 a_3} +
  {a_2\over a_1 a_3} + {a_3 \over a_1 a_2} \)
 Z_1 = 0 ~,\eqno(3.34a) $$
$$
{\hbar^2 \over 4}
a_i  {\pt B_2 \over \pt  a_i} +
4\pi^2 \hbar
  a_1 a_2 a_3 \( {a_1 \over a_2 a_3} +
  {a_2\over a_1 a_3} + {a_3 \over a_1 a_2} \)
 B_2
- {1 \over 4}  B_2 - 2\hbar^2 {\pt Z_1 \over \pt \ol\phi} -
4\pi^2 \hbar^2 \phi Z_1  = 0 ~,\eqno(3.34b)$$
$$
{\hbar^2 \over 4}
a_i  {\pt  B_1 \over \pt  a_i} -
4 \pi^2 \hbar
  a_1 a_2 a_3 \( {a_1 \over a_2 a_3} +
  {a_2\over a_1 a_3} + {a_3 \over a_1 a_2} \)
B_1
 + 2 \hbar^2 {\pt Z_1 \over \pt  \phi} +
 4\pi^2 \ol\phi Z_1  = 0~,
\eqno(3.34c) $$
$$
4\hbar^2
{\pt B_2 \over \pt \phi} + 8\pi^2 \ol
\phi B_2  - {3 \hbar^2 \over 4} a_i  {\pt Z_1 \over \pt a_i}
 + {3\hbar^2 \over 4 \sqrt{2}}
 Z_1 + 4\pi^2 \hbar
  a_1 a_2 a_3 \( {a_1 \over a_2 a_3} +
  {a_2\over a_1 a_3} + {a_3 \over a_1 a_2} \)
  Z_1 = 0 ~.\eqno(3.34d)$$
  We have used $a_i, i=1,2,3,$ to denote any operation with
  respect to $(a_1, a_2, a_3)$.
Making the   substitution
$B_1 = \tilde B_1 \exp ({-2\pi^2 \phi \bar \phi})$ ,
 $Z_1 ={\tilde Z_1 } \exp ({-2 \pi^2 \phi \bar \phi})$
 and $B_2 = \tilde B_2 \exp ({-2\pi^2 \phi
 \bar \phi})$ [25,40], the above four equations become
$$ 4 {\pt \tilde B_1 \over \pt \ol\phi} + {3 a_i \over 4}
{\pt \tilde Z_1 \over \pt a_i}
- {3 \over 4\sqrt{2}} \tilde Z_1
+
  4 \pi^2 \hbar^{-1} a_1 a_2 a_3 \( {a_1 \over a_2 a_3} +
  {a_2\over a_1 a_3} + {a_3 \over a_1 a_2} \)
\tilde Z_1 = 0 ~,\eqno (3.35a) $$
$$ 4  {\pt \tilde B_2 \over \pt \phi} -
{3 a_i \over 4} {\pt \tilde Z_1 \over \pt a_i}
+ {3 \over 4\sqrt{2}} \tilde Z_1 +
4\pi^2 \hbar^{-1}
  a_1 a_2 a_3 \( {a_1 \over a_2 a_3} +
  {a_2\over a_1 a_3} + {a_3 \over a_1 a_2} \)
  \tilde Z_1 = 0 ~,\eqno(3.35b) $$
$$ 2 {\pt \tilde Z_1 \over \pt \ol \phi} -
{3 \over 4} a_i {\pt \tilde B_2 \over \pt a_i}
- 4 \pi^2 \hbar^{-1}
  a_1 a_2 a_3 \( {a_1 \over a_2 a_3} +
  {a_2\over a_1 a_3} + {a_3 \over a_1 a_2} \)
 \tilde B_2 + {1 \over 4} a_1a_2 a_3 \tilde B_2 = 0 ~ \eqno(3.35c)$$
$$ 2 {\pt \tilde Z_1 \over \pt  \phi}
+ {1 \over 4} a_i {\pt \tilde B_1 \over \pt a_i} -
2\pi^2 \hbar^{-1}
  a_1 a_2 a_3 \( {a_1 \over a_2 a_3} +
  {a_2\over a_1 a_3} + {a_3 \over a_1 a_2} \)
\tilde B_1 = 0 ~.\eqno(3.35d) $$
We will now use (3.35a), (3.35d) and (3.35b), (3.35c) to
 eliminate $\tilde{B}_1$ or $\tilde{B}_2 $ to get a
differential equation for $\tilde{Z}_1$.
Let us take (3.35b), (3.35c). Applying
${\pt \over \pt  \phi}$ to (3.35c),
using  (3.35b) for ${\pt B_2  \over \pt \phi}$,
we obtain
$$  \eqalignno{ &2 {\pt \tilde Z_1 \over \pt \bar \phi \pt \phi}
 - {3 a_i \over 4^3} {\pt \over \pt a_i}
 \left(a_j {\pt \tilde Z_1 \over \pt a_j}\right) +
 {9 \over 4^3 \sqrt{2}} a_i {\pt \tilde Z_1 \over \pt a_i} \cr
&
+ \left[ 4 \pi^4 \hbar^{-2}
 \( a_1 a_2 a_3 \( {a_1 \over a_2 a_3} +
  {a_2\over a_1 a_3} + {a_3 \over a_1 a_2} \) \)^2 \right. \cr
& \left.  - \pi^2 \hbar^{-1}{{3 + \sqrt{2}} \over 4\sqrt{2}}
  a_1 a_2 a_3 \( {a_1 \over a_2 a_3} +
  {a_2\over a_1 a_3} + {a_3 \over a_1 a_2} \)
- {3 \over 4^3\sqrt{2}} \right] \tilde Z_1 = 0~.&
 (3.36)\cr} $$
We proceed similarly for the equations (3.35a), (3.35d)
but we notice that relatively to ref. [10,12,25],
the term in (3.35d)
linear in $\tilde{B}_1$  is absent.
This is due to the ordering we choose from the full
theory constraints. Had we chosen a different ordering
procedure
for the $S_A$ quantum constraint, as the one in  ref. [10,12,21] or
even as in [11] (see also ref. [25]),
the last column in table
1 for the $\beta$ row will have a bosonic
coefficient $B_1$.
However, it is easy to see  from the above equations
that the coefficients of the third, fifth and sixth
terms in (3.36) will differ from the ones in the
equation for $Z_1$ which we derive from the remaining equations
(3.35a), (3.35d).
Consistency than implies that
 $\tilde{Z}_1 = 0$.
Consequently, the equations for terms with
combinations of only $\beta$ and $\gamma$  involve just $B_1$ and
$C_1$. These are then as the ones in the case of
a Bianchi-IX with $\Lambda = 0$ and no supermatter [7].
The only possible solution of these equations
with respest to $a_1, a_2, a_3$ is  the
trivial one, i.e.,  $B_1 = C_1 = 0$. The equations corresponding
to $\ol\chi$ and  combinations of it with
$\beta$ or $\gamma$ would give, with $B_1 = C_1 = Z_1 = 0$,
the dependence of $B_2$ on $a_1,a_2,a_3,\phi$. This correspond to
$$B_2 = h(\ol\phi)
a_1a_2a_3 e^{-{{8\pi^2}\over \hbar}[a_1^2 + a_2^2 + a_3^2]}
e^{-2\pi^2\phi
\overline{\phi}}. \eqno(3.37)$$

This pattern repeats itself in  a similar way when we
consider the two groups  involving $D_1,D_2,E_1,Z_2$ and
$F_1,F_2,Z_3$. In particular, notice that from
the $\beta\ol\chi\gamma\gamma\gamma$ and $\beta\ol\chi\gamma$ terms in
$S_A \Psi = 0$ we get $E_1 = G_1 = H_1 = 0$ from
$Z_2 = D_2 = 0$, $Z_2 = D_1 = 0$, and $Z_3 = F_1 = 0$.

Hence, besides $A$ and $I_1$, only $B_2$ and
$F_2$ will be different from zero.
We can write than for the
solution of the constraints, using
the Ansatz (3.13),
$$\eqalignno{ \Psi &=
  f(\phi) e^{{{8\pi^2}\over \hbar}[a_1^2 + a_2^2 + a_3^2]}
\cr
&
+
 h(\ol\phi) a_1a_2a_3 e^{[-a_1^2
- a_2^2 - a_3^2]} e^{-2\pi^2 \phi
\overline{\phi}}
 \overline{\chi}_A\overline{\chi}^A \cr
&
+
 g(\phi) a_1a_2a_3 e^{{{8\pi^2}\over \hbar}[a_1^2 + a_2^2  + a_3^2]}
e^{-2\pi^2\phi
\overline{\phi}}
\beta_A\beta^A(\gamma_{BCD}\gamma^{BCD})^2 \cr
&+
  k(\ol\phi)  e^{{{8\pi^2}\over \hbar}[a_1^2 + a_2^2 + a_3^2]}
\overline{\chi}_A\overline{\chi}^A\beta_E\beta^E(\gamma_{BCD}\gamma^{BCD})^2.
& ~(3.38) \cr} $$

\bigskip

\medskip

{\bf IV. Conclusions, Discussions and Outlook}

In this paper we have studied the
quantization of a Bianchi type IX (diagonal) model
in N=1 supergravity in the presence of supermatter.
The supermatter content was constituted
by a scalar multiplet, i.e., a pair of complex
scalar fields together with their odd (anti-commuting)
spin-$\half$ fields partners. The corresponding
K\"ahler geometry  was chosen to be a two-dimensional
flat one.

Our approach can be characterized as twofold.
On the one hand, we applied directly the
quantum constraints of the full theory of
N=1 supergravity with supermatter [54]
subject to a (diagonal) Bianchi type-IX Ans\"atze for the
fields. On the other hand, we restricted ourselves
to a simple Lorentz invariant Ansatz for the
wave function of the universe, $\Psi$. In particular,
only the irreducible spin $\half, {3 \over 2}$ mode
components of the fermionic fields have been considered.
We then analysed two possible cases, namely when  the
scalar field dependent analytical potential
in the supermatter content was either arbitrary
or identically set to zero.

In the
former, our main (and perhaps, solely) conclusion
was that the adequacy of our Ansatz for $\Psi$ is
severly limited as far as chirality breaking,
mode mixing or non conservation
of fermion number (in the sense of [23,24])
are concerned. Following [23,24], it seems
paradoxical that for all the degrees of freedom    present
in the Bianchi type-IX with supermatter the constraints
imply $\Psi = 0$ for some simplifying assumptions.
These include either putting the analytical
potential term constant and different from zero
(as an effective cosmological constant) or choosing
some particular simple solutions for specific
bosonic amplitude coefficients in the Ansatz of $\Psi$.
In a more general setting, no easy way is apparent of
obtaining an analytical solution for the full set of equations.
In fact, exponential terms as $e^{\phi\ol\phi}$ lead to
serious difficulties.  However, the possibility
to have non-trivial states for $\Psi$ could not be ruled out
at cubic order in $\psi^A_i$ as in ref. [14,15,16]. In fact,
the presence of the matter degrees of freedom together
with the anisotropic gravitational ones and the
spin-$3\over 2$ modes of the gravitino allows one
to speculate on that possibility.

For $P(\phi)$, however, we found out that $\Psi$ had indeed
a very simple form. Namely, the only non zero
components of the wave function can be found in the sectors
with no fermions (bosonic) and in three other sectors, more precisely filled
with just the spin-$1 \over 2$
fermionic partners of the scalar field, another  filled with just the spin-$1
\over 2$ and
$3 \over 2$ mode components of the spatially homogeneous gravitino field
and finally one totally filled with spin-$1 \over 2$
fermionic partners of the scalar field as well as the
the spin-$1 \over 2$ and
$3 \over 2$ mode components of the gravitino field.
More precisely, we obtained
$$\eqalignno{ \Psi &=
  f(\phi) e^{-{{8\pi^2}\over \hbar}[a_1^2 + a_2^2 + a_3^2]}
\cr
&
+
 h(\ol\phi) a_1a_2a_3 e^{-{{8\pi^2}\over \hbar}[a_1^2
+ a_2^2 + a_3^2]} e^{-2\pi^2 \phi
\overline{\phi}}
 \overline{\chi}_A\overline{\chi}^A \cr
&
+
 g(\ol\phi) a_1a_2a_3 e^{{{8\pi^2}\over \hbar}[a_1^2 + a_2^2  + a_3^2]}
e^{-2\pi^2\phi
\overline{\phi}}
\beta_A\beta^A(\gamma_{BCD}\gamma^{BCD})^2 \cr
&+
  k(\ol\phi)  e^{{{8\pi^2}\over \hbar}[a_1^2 + a_2^2 + a_3^2]}
\overline{\chi}_A\overline{\chi}^A\beta_E\beta^E(\gamma_{BCD}\gamma^{BCD})^2.
 \cr} $$

The simple semi-classical form of the bosonic
amplitude coeficients in $\Psi$ above suggests that we might find
among them the Hartle-Hawking (no-boundary) state [45]
or the (ground) wormhole quantum state of the theory [46],
both of the form $P e^{-I}$ where $I$ is a certain
Euclidian action.
The wormhole state should correspond to an
asymptotic 4-Euclidian classical solution,
which is outwards to a 3-geometry,  required to be
regular at small 3-geometries (the interior boundary)
and to
die away rapidly at large 3-geometries [46].
For the Hartle-Hawking state, we require a regular
solution of the classical field equations with metric
(3.3) defined on the outer boundary [45].

The arbitrary functions $f, g, h, k$ of
$\phi, \ol\phi$ do not allow to conclude unambiguously
that in any of the fermionic
sectors the corresponding bosonic amplitudes
will die away for large 3-geometries and
$\phi, \ol\phi$ at  infinity.
Moreover, a wormhole state for a FRW case would have  the
form $prefactor \times e^{-3a^2 + 3a^2 \cosh(\rho)}$
where $\phi = \rho e^{i\theta}$ and one would expect a
simple generalization  of it to the Bianchi-IX case. Hence it seems that we
cannot find a
wormhole ground state. However, a similar issue for the case of a FRW model was
addressed recently (see ref. [62] for details) and where a wormhole basis may
still be constructed.

With regard to the Hartle-Hawking state,
the same arguments in [7] can be used to
show that we cannot identify such
type of solution in (3.37).
According to Graham and Luckock [8],
another
definition of homogeneity conditions for the
gravitino field {\it could} lead us to obtain
instead
$$\eqalignno{ \Psi &=
  f(\phi) e^{-{{8\pi^2}\over \hbar}[a_1^2 + a_2^2 + a_3^2
  + 2(a_2a_3 + a_1a_3 + a_1a_2)
  ]}
\cr
&
+
 h(\ol\phi) a_1a_2a_3 e^{-{{8\pi^2}\over \hbar}[a_1^2
+ a_2^2 + a_3^2 + 2(a_2a_3 + a_1a_3 + a_1a_2)]} e^{-2\pi^2 \phi
\overline{\phi}}
 \overline{\chi}_A\overline{\chi}^A \cr
&
+
 g(\phi) a_1a_2a_3 e^{{{8\pi^2}\over \hbar}
 [a_1^2 + a_2^2  + a_3^2 - 2(a_2a_3 + a_1a_3 + a_1a_2)]}
e^{-2\pi^2\phi
\overline{\phi}}
\beta_A\beta^A(\gamma_{BCD}\gamma^{BCD})^2 \cr
&+
 k(\ol\phi)  e^{{{8\pi^2}\over \hbar}[a_1^2 + a_2^2 + a_3^2
 - 2(a_2a_3 + a_1a_3 + a_1a_2)]}
\overline{\chi}_A\overline{\chi}^A\beta_E\beta^E(\gamma_{BCD}
\gamma^{BCD})^2,
 \cr} $$
where we  could identify the Hartle-Hawking state.

With respect to the models considered so far
within the more general N=1 supergravity
theory with supermatter [18,21,25,26,54], we
would expect our locally supersymmetric
(diagonal) Bianchi-IX model coupled to a
scalar supermultiplet to bear instead important
 differences. In fact, the models
in [18,21,25,26] were  FRW ones and
consequently the gravitino fields
were required to to be (severly) restricted to their
spin-$\half$ modes. However, the
presence now of anisotropic gravitational
degrees of freedom and hence of the spin-$3 \over 2$
modes of the gravitinos could play an important role.
Moreover, it would  bring our minisuperspace model closer
to the features of a full theory of N=1 supergravity
with supermatter in spite of the drastic
inhomogeneous modes truncation.
Our results corresponded to
``straightforward'' anisotropic generalization of
the previous works in the same line of research
[18,21,25,26]. Nevertheless,
we could not avoid some problems. In particular,
the absence of the Hartle-Hawking and wormhole states.
In the following, we would like to discuss (and somehow, to
speculate)
what may be the possible reasons
for that.

Firstly, we have obtained our differential
equations by applying the quantum constraints of the
full theory of N=1 supergravity with supermatter
subject a the Bianchi-IX (diagonal) Ansatz.
This has been used previously [6,7,14,15,16] and in agreement
with other approaches [34,35,36]. However, there are
some points which could be of some importance. We begin by
noticing that in this paper as well as in others following
the same approach [1,6,7,14,15,16],  contorsion terms
are absent in the full theory supersymmetry constraints.
This can be related to the inclusion of the Lorentz
constraints in the spatially integrated Hamiltonian through
the Lagrange multipliers $\omega^0_{AB}, \ol\omega^0_{A'B'}$.
However, in dimensional-reduced cases with supermatter
[10,11,12,18,21,25] contorsion cubic fermionic terms in the
gravitino field are present. In ref. [9,10] when no
supermatter was considered, a redefinition
of Lagrange multipliers was introduced in order
to obtain a simple form for the generators in the
Hamiltonian and their Dirac brackets; subsequetly the
contorsion cubic terms dissapeared.
It may be that when supermatter is present such
redefinitions cannot be straigthforwardly extended
from the pure N=1 supergravity case, either in the
full theory or from a dimensional reduction approach
(see [43] and sections 5,6 in [10]).

Secondly, we have used an overly restrictive Ansatz for
$\Psi$ in the sense that the Lorentz invariant sectors
have been constructed {\it without} the
irreducible spin components of the gravitational
degrees of freedom [23,24]. We are conscient of this fact
and our purpose was to obtain an indication
of what will be the picture in the
``new'' framework as explained by R. Graham and A. Csord\'as
[23,24].
Our arguments were as follows. Both the bosonic and
fermionic filled sectors amplitudes in [23,24] could be
obtained from  Lorentz invariant sectors
containing only  the irreducible spin components of the
gravitino field. Moreover, the  solutions for the
middle sectors (absent in the old framework) were allready
present in [8], using the old Ansatz for $\Psi$
together with a different homogeneity condition for gravitino.
The old framework uses a set of coupled first order
differential equations while the solutions for
middle states in [23,24]
(where the Hartle-Hawking state was properly identified)
depend on our ability to solve a (second-order) Wheeler-DeWitt
type of equation. Hence, we hoped that  the old
approach could still be of some utility as far as locally
supersymmetric models with supermatter are concerned.

It seems though from our analysis and discussion hereby
presented, that when supermatter is present one cannot
simply expect to follow a somewhat simple relation
between the old and new frameworks. Perhaps the approach
in [23,24] properly applied to our Bianchi-IX model could
be able to find out the
Hartle-Hawking states in {\it other} middle sectors.
However, the absence of a wormhole state is another
issue to be addressed [62]. Either a fundamental
piece in constructing the reduced theory
(following any of the possible approaches) has been
neglected or then we would have to conclude that
when  the more general
theory of N=1 supergravity with supermatter [54] is rightfully considered
 the wormhole picture may not be the ground state.

\bigskip

\bigskip

\noindent
{\bf ACKNOWLEDGEMENTS}

The author is  thankful to
A.D.Y. Cheng for many pleasant discussions
and for sharing his points of view and to
S.W. Hawking for helpful
 suggestions and conversations.
Discussions with  O. Obregon and R. Graham are also acnowledged.
The author   gratefully acknowledges
the Instituto de Fisica de la Universidad de Guanajuato, Mexico,
for supporting  part of  his 3 weeks visit
as well as
the support of
a Human Capital and Mobility (HCM)
Fellowship from the European Union (Contract ERBCHBICT930781).

\vskip .2 true in
\noindent
{\bf REFERENCES}

{\rm

\advance\leftskip by 4em
\parindent = -4em

[1] P.D. D'Eath, Phys.~Rev.~D {\bf 29}, 2199 (1984).

[2] A. Macias, O. Obr\'egon and M. Ryan,
Class. Quantum Grav. {\bf 4},  1477 (1987).

[3] O. Obr\'egon, J. Socorro and J. Benitez, Phys. Rev.
D {\bf 47}, 4471 (1993).

[4] J. Socorro, O. Obr\'egon and A. Macias,
Phys. Rev. D {\bf 45}, 2026 (1992).

[5] A. Macias, O. Obr\'egon and J. Socorro, Int. J. Mod.
Phys. {\bf A}8, 4291 (1993).

[6] P.D. D'Eath, S.W. Hawking and O. Obreg\'on, Phys.~Lett.~{\bf 300}B, 44
(1993).

[7] P.D. D'Eath, Phys.~Rev.~D {\bf 48}, 713 (1993).

[8] R. Graham and H. Luckock, Phys. Rev. D
{\bf 49}, R4981 (1994).

[9] P.D. D'Eath and D.I. Hughes, Phys.~Lett.~{\bf 214}B, 498 (1988).

[10] P.D. D'Eath and D.I. Hughes, Nucl.~Phys.~B {\bf 378}, 381 (1992).

[11] L.J. Alty, P.D. D'Eath and H.F. Dowker, Phys.~Rev.~D {\bf 46}, 4402
(1992).

[12] D.I. Hughes, Ph.D.~thesis, University of Cambridge (1990), unpublished.

[13] M. Asano, M. Tanimoto and N. Yoshino, Phys.~Lett.~{\bf 314}B, 303 (1993).

[14] P.D. D'Eath, Phys. Lett. B{\bf 320}, 20 (1994).

[15]  A.D.Y. Cheng, P.D. D'Eath and
P.R.L.V. Moniz, Phys. Rev. D{\bf 49} (1994) 5246.

[16]  A.D.Y. Cheng, P.D. D'Eath and
P.R.L.V. Moniz,
{\rm Gravitation and Cosmology}
{\bf 1} (1995) 1

[17] A.D.Y. Cheng, P.D. D'Eath and
P.R.L.V. Moniz,
{}~DAMTP-Report February R94/13, submitted to
Physical Review D.

[18]  A.D.Y. Cheng, P.D. D'Eath and
P.R.L.V. Moniz,
{\rm Gravitation and Cosmology - Proceedings}
{\bf 1} (1995) 12

[19] S. Carroll, D. Freedman, M. Ortiz and
D. Page, Nuc. Phys. B{\bf 423}, 3405 (1994).

[20] S. Carroll, D. Freedman, M. Ortiz and
D. Page, {\it Bosonic
physical states in N=1 supergravity?}
in: Procedings of the MG7 Meeting in General
Relativity, Stanford, July 1994 (World Scientific),
gr-qc 9410005.

[21] A.D.Y. Cheng, P.D. D'Eath and
P.R.L.V. Moniz,
Class. Quantum Grav., to appear; {\it Quantization
of a FRW model in N=1 supergravity
with gauged supermatter}
in:
Proceedings of the First Mexican School in Gravitation
and Mathematical Physics,
Guanajuato, December 12-16, 1994 (9503009).

[22] H. Luckock and C. Oliwa, Sidney University Report,
(gr-qc 9412028), accepted for publication in
Phys. Rev. D.; C. Oliwa, M.Sc. thesis, University
of Sidney (1994), unpublished.

[23] R. Graham and A. Csord\'as, {\it
Nontrivial fermion states in supersymmetric minisuperspace},
in:
Proceedings of the First Mexican School in Gravitation
and Mathematical Physics,
Guanajuato, Mexico, December 12-16, 1994 (9503054 ).

[24]  R. Graham and A. Csord\'as, Phys. Rev. Lett. {\bf 74} (1995) 4926.

[25]  \&
Int. J. Mod. Phys. {\bf D4}, No.2 April (1995) - to appear.

[26]   A.D.Y. Cheng, unpublished report.

[27] R. Graham, Phys.~Rev.~Lett.~{\bf 67}, 1381 (1991).

[28]  R. Graham and J. Bene, Phys. Lett. B {\bf 302}, 183 (1993).

[29] R. Graham, Phys. Rev. D {\bf 48}, 1602 (1993).

[30] R. Graham and J. Bene,  Phys. Rev. D{\bf 49}, 799 (1994).

[31] R. Graham and H. Luckock, Phys. Rev. D {\bf 49}, 2786 (1994).

[32] T. Jacobson, Class. Quantum Grav. {\bf 5}, 923, (1988).

[33]  T. Sano and J. Shiraishi, Nucl. Phys. B{\bf 410}, 423, (1993).

[34] O. Obr\'egon, J. Pullin and M. Ryan, Phys. Rev. D {\bf 48},
5642 (1993).

[35] R. Capovilla and J. Guven, Class. Quantum Grav.
{\bf 11}, 1961 (1994).

[36] R. Capovilla and O. Obregon, Phys. Rev. D{\bf 49} (1994) 6562.

[37] H.-J. Matschull, Class. Quantum Grav. {\bf 11} (1994) 2395.

[38]    H.-J. Matschull and H. Nicolai, Jour. Geom. Phys.
{\bf 11}, 15 (1993); Nucl. Phys. B {\bf 411}, 609 (1994);
B. de Wit, H.-J. Matschull and H. Nicolai,  Phys.
Lett. B {\bf 318}, 115 (1993).

[39] T. Sano, hep-th 9211103; T. Sano and J. Shiraishi,
Nuc. Phys. {\bf B}410 (193) 423.

[40] H. Kunimoto and T. Sano,
Int. J. Mod. Phys. {\bf D1} (1993) 559.

[41]  P. V. Moniz, {\it  Supersymmetric Quantum  Cosmology},
work in preparation.

[42] C. Teitelboim, Phys.~Rev.~Lett.~{\bf 38}, 1106 (1977).

[43] M. Pilati, Nuc. Phys. B {\bf 132}, 138 (1978).

[44]  M. Ryan Jr. and L. Shepley, ``Homogeneous
Relativistic Cosmologies'' (PUP, Princeton 1972).

[46] S.W. Hawking and D.N. Page, Phys.~Rev.~D {\bf 42}, 2655 (1990).

[45] J.B. Hartle and S.W. Hawking, Phys.~Rev.~D {\bf 28}, 2960 (1983).

[47]   C. Teitelboim, Phys. Rev. D {\bf 25},  3159 (1982)

[48]   A.D.Y. Cheng, private communication.

[49]   R. Graham, private communication.

[50]   P. van Nieuwenhuizen, Phys.~Rep. {\bf 68} (1981) 189;
D. Freedman and A. Das, Nuc. Phys. B {\bf 120}, 221 (1977).

[51]  S. Ferrara and P. van Nieuwenhuizen,
Phys. Rev. Lett. {\bf 37}, 1669 (1976).

[52]  D. Freedman,
Phys. Rev. Lett. {\bf 38}, 105 (1977).

[53] \&, work in progress.

[54]  J. Wess and J. Bagger, {\it Supersymmetry and Supergravity},
2nd.~ed. (Princeton University Press, 1992).

[55] A. Das. M. Fishler and M. Rocek,  Phys.~Lett.~B {\bf 69}, 186 (1977).

[56] O. Obr\'egon, IFUG preprint-1995.

[57] O. Obr\'egon, private communication.

[58] J.E. Nelson and C. Teitelboim, Ann.~Phys. (N.Y.) {\bf 116} (1978)
86.

[59] P.D. D'Eath and J.J. Halliwell, Phys.~Rev.~D {\bf 35} (1987) 1100.

[60] E. Witten and J. Bagger, Phys.~Lett.~B {\bf 115} (1982).

[61]  L.D. Faddeev and A.A. Slavnov, {\it Gauge Fields} (Benjamin,
Reading, Mass., 1980).

[62] P.V. Moniz, {\it Is there a problem with quantum
wormholes in N=1 supergravity}, essay-DAMTP R95/19;
{\it The case of the missing wormhole case},
talk present in the 6th Moskow Quantum Gravity Seminar.

[63] P.V. Moniz, {\it Locally supersymmetric
FRW model with Yang-Mills fields}, in preparation.

\ \ \ }

\bye